\documentclass[preprint,12pt]{elsarticle}

\usepackage{graphicx}
\graphicspath{{Figures/}}
\usepackage{subfigure}
\usepackage{amssymb}
\usepackage{amsmath}
\usepackage{amsfonts}

\usepackage{lineno}

\usepackage{xcolor}

\journal{Journal Name}

\begin{document}

\begin{frontmatter}


\title{Reduced-order modeling of advection-dominated systems with recurrent neural networks and convolutional autoencoders}



\author[label1]{Romit Maulik}
\author[label1]{Bethany Lusch}
\author[label2,label1]{Prasanna Balaprakash}

\address[label1]{Argonne Leadership Computing Facility, Argonne National Laboratory, Lemont - IL, 60439}
\address[label2]{Mathematics \& Computer Science Division, Argonne National Laboratory, Lemont - IL, 60439}

\begin{abstract}
A common strategy for the dimensionality reduction of nonlinear partial differential equations relies on the use of the proper orthogonal decomposition (POD) to identify a reduced subspace and the Galerkin projection for evolving dynamics in this reduced space. However, advection-dominated PDEs are represented poorly by this methodology since the process of truncation discards important interactions between higher-order modes during time evolution. In this study, we demonstrate that an encoding using convolutional autoencoders (CAEs) followed by a reduced-space time evolution by recurrent neural networks overcomes this limitation effectively. We demonstrate that a truncated system of only two latent-space dimensions can reproduce a sharp advecting shock profile for the viscous Burgers equation with very low viscosities, and a twelve-dimensional latent space can recreate the evolution of the inviscid shallow water equations. Additionally, the proposed framework is extended to a parametric reduced-order model by directly embedding parametric information into the latent space to detect trends in system evolution. Our results show that these advection-dominated systems are more amenable to low-dimensional encoding and time evolution by a CAE and recurrent neural network combination than the POD Galerkin technique.
\end{abstract}


\end{frontmatter}


\section{Introduction}
\label{S:1}

High-fidelity simulations of systems characterized by nonlinear partial differential equations (PDEs) represent large compute costs and are prohibitive for decision-making tasks for many fast-query applications. In order to reduce costs, there has recently been significant interest in the reduced-order modeling (ROM) of such systems \cite{carlberg2011efficient,wang2012proper,san2015principal,ballarin2015supremizer,san2018extreme,wang2019non,choi2019space}. As such, this field finds extensive application in control \cite{proctor2016dynamic,peitz2019multiobjective,noack2011reduced,rowley2017model}, multi-fidelity optimization \cite{peherstorfer2016optimal}, uncertainty quantification \cite{sapsis2013statistically,zahr2018efficient} and data-assimilation \cite{arcucci2019optimal} among others. However, ROMs are limited in how they handle nonlinear dependence and perform poorly for complex physical phenomena, which are inherently multiscale in space and time \cite{wells2017evolve,xie2018data,san2018neural,san2019artificial}. Researchers continue to search for efficient and reliable ROM techniques for such transient nonlinear systems \cite{hess2019localized,kramer2019nonlinear,swischuk2019projection,hamzi2019local,rahman2019dynamic,wang2019non}. The identification of a reduced basis to ensure a compressed representation that is minimally \emph{lossy} is a core component of most ROM development strategies (some examples include \cite{san2014basis,korda2018data,kalb2007intrinsic}). Once this basis is identified, we need a cost-effective strategy for accurate nonlinear dynamical system evolution to reproduce the full-order spatiotemporal complexity of the problem in the reduced basis. For example, we could use intrusive methods (which project the governing equations onto the reduced-basis), as seen in \cite{kalashnikova2010stability,mohebujjaman2019physically}, which use a Galerkin projection or \cite{carlberg2011efficient,xiao2013non,fang2013non}, which use the Petrov-Galerkin method (see \cite{carlberg2017galerkin} for the comparison of these two methods). Finally, reconstruction of the compressed representation is required for full-order space post-processing and visualization. In this study, we utilize convolutional autoencoders (CAE) and long short-term memory neural networks (LSTM) \cite{hochreiter1997long} for parametric ROMs of advection-dominated and inviscid systems. The former are used to identify reduced-representations of the full-order fields and the latter are used for the temporal evolution of these compressed representations. LSTMs have recently become popular for the non-intrusive characterization of dynamical systems \cite{vlachas2018data,ahmed2019long,maulik2019time,mohan2019compressed,mohan2018deep,wang2019recurrent} although most such studies perform latent space embedding through the use of linear embeddings such as the proper orthogonal decomposition (POD). Additionally, we propose a parametric extension of the CAE-LSTM for exploring parametric search spaces through training on multiple offline simulations. In contrast with studies outlined in \cite{lusch2018deep,erichson2019physics}, we deploy our framework on problems requiring shock capturing mechanisms as well as for fully inviscid system simulations. Ref. \cite{lee2020model} also uses a CAE to nonlinearly embed states, but solves the governing equations on the
nonlinear manifold defined by the autoencoder. We reduce computation by evolving with an LSTM network instead. An important difference with another similar study outlined in \cite{gonzalez2018learning} is that our framework allows for the explicit embedding of control parameters such as viscosity or a parameterization of the initial condition into the LSTM. This allows for the independent training of our CAE and LSTM neural networks. A similar implementation has been demonstrated in \cite{xu2019multi} where another neural network is used to link global parameters with the latent space representations. We simplify this by directly embedding the parameters into the latent space. In \citep{maulik2020latent}, the latent space was obtained using a CAE followed by which a Gaussian process regressor was utilized for a continous representation of the temporal evolution of the state. However, this method is limited when the number of snapshots is very large due to the inherent limitations of most Gaussian process regression techniques.

Our forward models for the purpose of data generation and subsequent testing are given by parametric partial differential equations. For the rest of this article, we shall represent a generic partial differential equation using the following notation:
\begin{linenomath*}
\begin{align}
\label{gen1}
\dot{\mathbf{q}}(\mathbf{x},t,\mathbf{p}) + \mathcal{N}[\mathbf{q}(\mathbf{x},t,\mathbf{p})] + \mathcal{L}[\mathbf{q}(\mathbf{x},t,\mathbf{p}); \mathbf{p}] = 0, \quad (\mathbf{x},t,\mathbf{p}) \in \Omega \times \mathcal{T} \times \mathcal{P},
\end{align}
\end{linenomath*}
where $\Omega \subset \mathbb{R}^i, \mathcal{T} = [0,T], \mathcal{P} \subset \mathbb{R}^p$, and $\mathcal{N}$, $\mathcal{L}$ are non-linear and linear operators, respectively. Our system is characterized by a solution field $\mathbf{q} : \Omega \times \mathcal{T} \times \mathcal{P} \rightarrow \mathbb{R}^d$ and appropriately chosen initial and boundary conditions, where $i$ is the number of spatial dimensions, $d$ is the number of dependent variables of the PDE, and $p$ is the number of control parameters in the problem. We assume that our system of equations can be solved in space-time on a discrete grid resulting in the following system of parameterized ODEs:
\begin{linenomath*}
\begin{align}
\dot{\mathbf{q}_h}(t,\mathbf{p}) + \mathbf{N}_{h}[\mathbf{q}_h(t,\mathbf{p})] + \mathbf{L}_h[\mathbf{q}_h(t,\mathbf{p}); \mathbf{p}] = 0  \quad(t,\mathbf{p}) \in \mathcal{T} \times \mathcal{P},
\end{align}
\end{linenomath*}
where $\mathbf{q}_h : \mathcal{T} \times \mathcal{P} \rightarrow \mathbb{R}^{N_h}$ is a discrete solution and $N_h$ is the number of spatial degrees of freedom. In this problem, our goal is to bypass the solution of Equation \ref{gen1} by constructing a compression manifold and a time advancement technique on this manifold solely from training data. Such ROMs hold great promise for characterizing the spatiotemporal dynamics of systems for which observations may be available, but little knowledge of the governing equations exist. 

{To summarize, the contributions of this article are:
\begin{itemize}
    \item We propose a deep learning based emulation strategy for nonlinear partial differential equations.
    \item We introduce a convolutional autoencoder architecture that obtains nonlinear embeddings with high compression ratios.
    \item We propose the use of long short-term memory networks for the evolution of the state in embedded space.
    \item We make our emulator parametric by passing the parameters into the latent space. This allows generalization across, for example, a range of viscosities.
    \item We demonstrate the performance of the proposed formulation for non-intrusive modeling of advection dominated physics obtained from the viscous Burgers and inviscid shallow water equations.
\end{itemize}}

\section{Proper orthogonal decomposition}
\label{S:2}

In this section, we review the POD technique for the construction of a reduced basis \cite{kosambi1943statistics,berkooz1993proper}. The interested reader may also find an excellent explanation of POD and its relationship with other dimension-reduction techniques in \cite{taira2019modal}. The POD procedure is tasked with identifying a space
\begin{linenomath*}
\begin{align}
\mathbf{X}^{f}=\operatorname{span}\left\{\boldsymbol{\vartheta}^{1}, \dots, \boldsymbol{\vartheta}^{f}\right\},
\end{align}
\end{linenomath*}
which approximates snapshots optimally with respect to the $L^2-$norm. The process of $\boldsymbol{\vartheta}$ generation commences with the collection of snapshots in the \emph{snapshot matrix}
\begin{linenomath*}
\begin{align}
\mathbf{S} = [\begin{array}{c|c|c|c}{\hat{\mathbf{q}}^{1}_h} & {\hat{\mathbf{q}}^{2}_h} & {\cdots} & {\hat{\mathbf{q}}^{N_{s}}_h}\end{array}] \in \mathbb{R}^{N_{h} \times N_{s}},
\end{align}
\end{linenomath*}
where $N_s$ is the number of snapshots, and $\hat{\mathbf{q}}^i_h \in \mathbb{R}^{N_h}$ corresponds to an individual snapshot in time of the discrete solution domain. Our POD bases can then be extracted efficiently through the method of snapshots where we solve the eigenvalue problem on the correlation matrix $\mathbf{C} = \mathbf{S}^T \mathbf{S} \in \mathbb{R}^{N_s \times N_s}$. Then
\begin{linenomath*}
\begin{align}
\begin{gathered}
\mathbf{C} \mathbf{W} = \mathbf{W} \Lambda,
\end{gathered}
\end{align}
\end{linenomath*}
where $\Lambda = \operatorname{diag}\left\{\lambda_{1}, \lambda_{2}, \cdots, \lambda_{N_{s}}\right\} \in \mathbb{R}^{N_{s} \times N_{s}}$ is the diagonal matrix of eigenvalues and $\mathbf{W} \in \mathbb{R}^{N_{s} \times N_{s}}$ is the eigenvector matrix. Our POD basis matrix can then be obtained by
\begin{linenomath*}
\begin{align}
\begin{gathered}
\boldsymbol{\vartheta} = \mathbf{S} \mathbf{W} \in \mathbb{R}^{N_h \times N_s}.
\end{gathered}
\end{align}
\end{linenomath*}
In practice a reduced basis $\boldsymbol{\psi} \in \mathbb{R}^{N_h \times N_r}$ is built by choosing the first $N_r$ columns of $\boldsymbol{\vartheta}$ for the purpose of efficient ROMs, where $N_r \ll N_s$. This reduced basis spans a space given by
\begin{linenomath*}
\begin{align}
\mathbf{X}^{r}=\operatorname{span}\left\{\boldsymbol{\psi}^{1}, \dots, \boldsymbol{\psi}^{N_r}\right\}.
\end{align}
\end{linenomath*}
The coefficients of this reduced basis (which capture the underlying temporal effects) may be extracted as
\begin{linenomath*}
\begin{align}
\begin{gathered}
\mathbf{A} = \boldsymbol{\psi}^{T} \mathbf{S} \in \mathbb{R}^{N_r \times N_s}.
\end{gathered}
\end{align}
\end{linenomath*}
The POD approximation of our solution is then obtained via
\begin{linenomath*}
\begin{align}
\hat{\mathbf{S}} =  [\begin{array}{c|c|c|c}{\tilde{\mathbf{q}}^{1}_h} & {\tilde{\mathbf{q}}^{2}_h} & {\cdots} & {\tilde{\mathbf{q}}^{N_{s}}_h}\end{array}] \approx \boldsymbol{\psi} \mathbf{A} \in \mathbb{R}^{N_h \times N_s},
\end{align}
\end{linenomath*}
where $\tilde{\mathbf{q}}_h^i \in \mathbb{R}^{N_h}$ corresponds to the POD approximation to $\hat{\mathbf{q}}_h^i$. The optimal nature of reconstruction may be understood by defining the relative projection error
\begin{linenomath*}
\begin{align}
\frac{\sum_{i=1}^{N_{s}}\left\|\hat{\mathbf{q}}^i_h-\tilde{\mathbf{q}}^i_h \right\|_{\mathbb{R}^{N_{h}}}^{2}}{\sum_{i=1}^{N_{s}}\left\|\hat{\mathbf{q}}^i_h\right\|_{\mathbb{R}^{N_{h}}}^{2}}=\frac{\sum_{i=N_r+1}^{N_{s}} \lambda_{i}^{2}}{\sum_{i=1}^{N_{s}} \lambda_{i}^{2}},
\end{align}
\end{linenomath*}
which exhibits that with increasing retention of POD bases, increasing reconstruction accuracy may be obtained. We remark that for dimension $d>1$, the solution variables may be stacked to obtain this set of bases that are utilized for the reduction of each PDE within the coupled system. Another approach may be to obtain reduced bases for each dependent variable within the coupled system and evolve each PDE on a different manifold. Each dependent variable is projected onto bases constructed from its snapshots alone. This affects the computation of $\mathcal{N}$ for computing the updates for each dimension in $\mathbf{q}$. In practice, this operation manifests itself in the concatenation of reduced bases to obtain one linear operation for reconstruction of all field quantities.

\subsection{The POD Galerkin projection}

The POD basis may be leveraged for a Galerkin projection of each partial differential equation forming the coupled system onto its corresponding reduced basis. We start by revisiting Equation (\ref{gen1}) written in the form of an evolution equation for fluctuation components i.e.,
\begin{linenomath*}
\begin{align}
\dot{\hat{\mathbf{q}}}_h(\mathbf{x},t,\mathbf{p}) + \mathcal{N}_h[\hat{\mathbf{q}}_h(\mathbf{x},t,\mathbf{p})] + \mathcal{L}_h[\hat{\mathbf{q}}_h(\mathbf{x},t,\mathbf{p}); \mathbf{p}] = 0,
\end{align}
\end{linenomath*}
which can be expressed in the reduced basis as 
\begin{linenomath*}
\begin{align}
\boldsymbol{\psi} \dot{\mathbf{q}_r}(t,\mathbf{p}) + \mathcal{N}_h[\boldsymbol{\psi} \mathbf{q}_r(t,\mathbf{p})] + \mathcal{L}_h[\boldsymbol{\psi} \mathbf{q}_r(t,\mathbf{p}); \mathbf{p}] = 0,
\end{align}
\end{linenomath*}
where $\mathbf{q}_r \in \mathbb{R}^{N_r}$ corresponds to the temporal coefficients at one time instant of the system evolution (i.e., equivalent to a particular column of $\mathbf{A}$). The orthogonal nature of the reduced basis can be leveraged to obtain
\begin{linenomath*}
\begin{align}
\dot{\mathbf{q}_r}(t,\mathbf{p}) + \mathcal{N}_h[\boldsymbol{\psi} \mathbf{q}_r(t,\mathbf{p})] + \mathcal{L}_r[\mathbf{q}_r(t,\mathbf{p}); \mathbf{p}] = 0,
\end{align}
\end{linenomath*}
where $\mathcal{L}_r$ is a precomputed Laplacian operator in reduced space. This equation is denoted the POD Galerkin-projection formulation (POD-GP). We have assumed that the residuals generated by the truncated representation of the full-order model are orthogonal to the reduced basis. A significant source of error in the forward evolution of this system of equations is due to the absence of higher-basis nonlinear interactions as shown in in Section \ref{SS:6}. Also, POD-GP essentially consists of $N_r$ coupled ODEs and is solved by a standard fourth-order accurate Runge-Kutta method. The reduced degrees of freedom lead to very efficient forward solves of the problem even though accuracy is limited. This transformed problem has initial conditions given by
\begin{linenomath*}
\begin{align}
\mathbf{q}_r(t=0)=\left(\boldsymbol{\psi}^T \hat{\mathbf{q}}_h(t=0) \right).
\end{align}
\end{linenomath*}

\section{Deep neural networks}

In the following section, we introduce our deep neural network architectures for establishing a viable emulation strategy for data obtained from nonlinear partial differential equations.

\subsection{Convolutional autoencoders}
\label{S:3}
Autoencoders are neural networks that learn a new representation of the input data, usually with lower dimensionality. The initial layers, called the \emph{encoder}, map the input $\mathbf{x}\in \mathbb{R}^m$ to a new representation $\mathbf{z} \in \mathbb{R}^k$ with $k << m$. The remaining layers, called the \emph{decoder}, map $\mathbf{z}$ back to $\mathbb{R}^m$ with the goal of reconstructing $\mathbf{x}$. The objective is to minimize the reconstruction error. Autoencoders are unsupervised; the data $\mathbf{x}$ is given, but the representation $\mathbf{z}$ must be learned.

More specifically, we use autoencoders that have some convolutional layers. In a convolutional layer, instead of learning a matrix that connects all $m$ neurons of layer's input to all $n$ neurons of the layer's output, we learn a set of filters.  Each filter $\mathbf{f_i}$ is convolved with patches of the layer's input. Suppose a 1-d convolutional layer has filters of length $m_{f_i}$. Then each of the layer's output neurons corresponding to filter $\mathbf{f_i}$ is connected to a patch of $m_{f_i}$ of the layer's input neurons. In particular, a 1-d convolution of filter $\mathbf{f}$ and patch $\mathbf{p}$ is defined as $\mathbf{f} \ast \mathbf{p} = \sum_j f_j p_j$ (For neural networks, convolutions are usually technically implemented as cross-correlations). Then, for a typical 1-d convolutional layer, the layer's output neuron $y_{ij} = \varphi (\mathbf{f_i} \ast \mathbf{p_j} +B_{i})$ where $\varphi$ is an activation function, and $B_i$ are the entries of a bias term. As $j$ increases, patches are shifted by stride $s$. For example, a 1-d convolutional layer with a filter $f_0$ of length $m_f = 3$ and stride $s=1$ could be defined so that $y_{0j}$ involves the convolution of $f_0$ and inputs $j-1, j$, and $j+1$. To calculate the convolution, it is common to add zeros around the inputs to a layer, which is called \emph{zero padding}. In the decoder, we use deconvolutional layers to return to the original dimension. These layers upsample with nearest-neighbor interpolation.

Two-dimensional convolutions are defined similarly, but each filter and each patch are two-dimensional. A 2-d convolution sums over both dimensions, and patches are shifted both ways. For a typical 2-d convolutional layer, the output neuron $y_{ijk} = \varphi (\mathbf{f_i} \ast \mathbf{p_{jk}} +B_{i})$. Input data can also have a ``channel'' dimension, such as RGB for images. The convolutional operator sums over channel dimensions, but each patch contains all of the channels. The filters remain the same size as patches, so they can have different weights for different channels. It is common to follow a convolutional layer with a \emph{pooling} layer, which outputs a sub-sampled version of the input. In this paper, we specifically use max-pooling layers. Each output of a max-pooling layer is connected to a patch of the input, and it returns the maximum value in the patch. 

\subsection{Long short-term memory networks}
\label{S:4}

The LSTM network is a specialization of the recurrent neural network and was introduced to consider time-delayed processes where events further back in the past may potentially affect predictions for the current location in the sequence. The basic equations of the LSTM in our context for an arbitrary input variable $\mathbf{a}$ are given by
\begin{linenomath*}
\begin{align}
\begin{split}
\text{input gate: }& \boldsymbol{G}_{i}=\boldsymbol{\varphi}_{S} \circ \mathcal{F}_{i}^{N_{c}}(\mathbf{a}), \\
\text{forget gate: }& \boldsymbol{G}_{f}=\boldsymbol{\varphi}_{S} \circ \mathcal{F}_{f}^{N_{c}}(\mathbf{a}), \\
\text{output gate: }& \boldsymbol{G}_{o}=\boldsymbol{\varphi}_{S} \circ \mathcal{F}_{o}^{N_{c}}(\mathbf{a}), \\
\text{internal state: }& \boldsymbol{s}_{t}=\boldsymbol{G}_{f} \odot \boldsymbol{s}_{t-1}+\boldsymbol{G}_{i} \odot\left(\boldsymbol{\varphi}_{T} \circ \mathcal{F}_{\mathbf{a}}^{N_{c}}(\mathbf{a})\right), \\
\text{output: }& \mathbf{h}_t = \boldsymbol{G}_{o} \circ \boldsymbol{\varphi}_{T}\left(\boldsymbol{s}_{t}\right),
\end{split}
\end{align}
\end{linenomath*}
where $\mathbf{a}$ is a vector of inputs comprising a snapshot of information in time. Within this study, this vector is generally the encoded representation after either the POD or CAE embedding. Also, $\boldsymbol{\varphi}_{S}$ and $\boldsymbol{\varphi}_{T}$ refer to tangent sigmoid and tangent hyperbolic activation functions respectively, $N_c$ is the number of hidden layer units in the LSTM network. Here, $\mathcal{F}^{n}$ refers to a linear operation given by a matrix multiplication and subsequent bias addition i.e,
\begin{linenomath*}
\begin{align}
\mathcal{F}^{n}(\boldsymbol{x})=\boldsymbol{W} \boldsymbol{x}+\boldsymbol{B},
\end{align}
\end{linenomath*}
where $\boldsymbol{W} \in \mathbb{R}^{n \times m}$ and $\boldsymbol{B} \in \mathbb{R}^{n}$ for $\mathbf{x} \in \mathbb{R}^m$ and where $\mathbf{a} \odot \mathbf{b}$ refers to a Hadamard product of two vectors. The LSTM implementation is used to advance $\mathbf{a}$ as a function of time. The LSTM network's primary utility is the ability to control information flow through time with the use of the gating mechanisms. A quantity that preserves information of past inputs and predictions is the internal state $\mathbf{s}_t$ which is updated using the result of the input and forget gates every time the LSTM operations are performed. A greater value of the forget gate (post sigmoidal activation), allows for a greater preservation of past state information through the sequential inference of the LSTM, whereas a smaller value suppresses the influence of the past. Details of our LSTM deployments for the different experiments utilized in this article are provided in Section \ref{S:5}. 

\subsection{Combining CAE and LSTM for surrogate modeling}

{Our data-driven emulation strategy shall rely on the use of CAE for dimensionality reduction and LSTM for latent space temporal evolution of the state. The benefit of this formulation, in comparison to POD-GP, is the improved compression ratios obtained by the nonlinear embedding of the CAE and the equation-free evolution of the state using the LSTM. The basic schematic for this formulation is shown in Figure \ref{1D_Schematic} where a one-dimensional field is compressed to a low-dimensional latent space and then evolved non-intrusively. Our training framework is \emph{separate} in that the snapshot data of the flow-field is first used to obtain a low dimensional embedding before a data-driven time-series forecast technique is used for evolving the state in this space. As mentioned previously, this is in contrast to previous studies where latent space embedding and temporal evolution have been performed in a simultaneous optimization \cite{lusch2018deep,gonzalez2018learning,erichson2019physics}. We pursue this direction for our emulation strategy to allow for greater flexibility in modeling the evolution of the latent space. In particular, the choice for a novel state evolution mechanism will not require retraining a nonlinear embedding. In addition, uneven samples of snapshot data (for instance when new snapshots become available at arbitrary locations in time) may be deployed with time-series methods that are customized for irregular data without retraining an embedding \cite{rubanova2019latent}. The deployment of a concurrent optimization for an embedding and a time series forecast strategy usually relies on the construction of a loss-function that penalizes reconstruction and forecast accuracy together. The joint optimization can result in slower training and requires deciding how to weight the two loss functions. Specific details of the CAE and LSTM combinations for our test cases shall be described in Section \ref{S:5}.}

\section{Experiments}
\label{S:5}

In the following, we introduce the two representative problems used to assess the proposed framework. We demonstrate framework performance for the viscous Burgers equation, which is characterized by an advecting shock and the conservative inviscid shallow water equations with varying initial conditions. While the first problem requires that our framework is able to capture the advection of a shock profile accurately in time, the second problem requires interpolation in initial condition space. These varying initial conditions are given by different locations of a Gaussian blob at the starting time. The specific details of the ML framework used in the following experiments may be found in our supporting source code available at \texttt{https://github.com/Romit-Maulik/CAE\_LSTM\_ROMS}.

\subsection{Burgers}
\label{SS:6}

Our first problem is given by the one-dimensional viscous Burgers' equation with Dirichlet boundary conditions which can be represented as
\begin{linenomath*}
\begin{align}
\begin{gathered}
\label{gen3}
\dot{u} + u\frac{\partial u}{\partial x} = \nu \frac{\partial^2 u}{\partial x^2}, \\
u(x,0) = u_0, \quad x \in [0,L], \quad u(0,t) = u(L,t) = 0.
\end{gathered}
\end{align}
\end{linenomath*}
It is well known that the above equation is capable of generating discontinuous solutions even if initial conditions are smooth and $\nu$ is sufficiently small due to advection-dominated behavior. We specifically consider the initial condition
\begin{linenomath*}
\begin{align}
u(x, 0) &=\frac{x}{1+\sqrt{\frac{1}{t_{0}}} \exp \left(R e \frac{x^{2}}{4}\right)}, 
\end{align}
\end{linenomath*}
and we set $L=1$ and maximum time $t_{max}=2$. An analytical solution exists and is given by
\begin{linenomath*}
\begin{align}
\label{Burgers_Sol}
u(x, t)=\frac{\frac{x}{t+1}}{1+\sqrt{\frac{t+1}{t_{0}}} \exp \left(R e \frac{x^{2}}{4 t+4}\right)},
\end{align}
\end{linenomath*}
where $t_0=\text{exp}(Re/8)$ and $Re = 1/\nu$.

\subsubsection{Convolutional autoencoder} \label{Burgers_CAE}

\begin{figure}
    \centering
    \fbox{\includegraphics[width=0.95\textwidth]{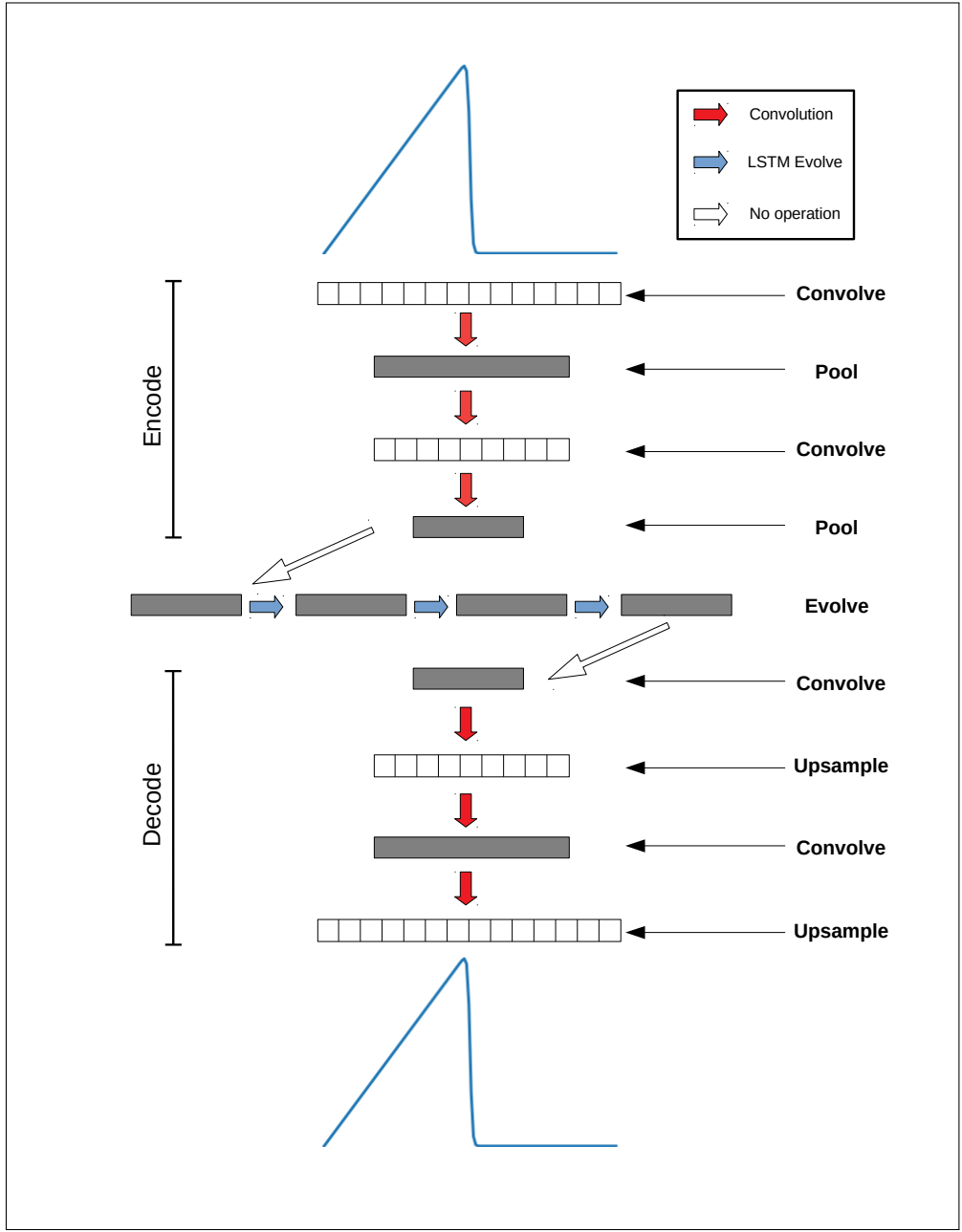}}
    \caption{A schematic of the one-dimensional CAE-LSTM for the viscous Burgers equation. The nonlinear autoencoder embeds the data into latent space, and then the recurrent network can be used for time-series advancement.}
    \label{1D_Schematic}
\end{figure}

We proceed by detailing the architecture of our CAE for effective compression of the full-order solution field. We use a one-dimensional convolutional framework with multiple strided filters to obtain a low-dimensional representation of the solution field. Figure \ref{1D_Schematic} is a schematic of the architecture. We utilize several pairs of convolutional and max-pooling layers to reduce dimensionality of the input image to a size of \emph{solely} two degrees of freedom in the encoded space. Following this, the two-dimensional state is convolved and upsampled several times to return to the dimensionality of the full-order field. Each layer consists of rectified linear (ReLU) activations and utilizes a zero-padding at the edges of the domain for the purpose of convolution. The dynamics studied in this test case are not critically affected by the absence of accurate padding at the boundaries. Our network is trained by using a standard mean-squared error loss with a batch size of 10, a learning rate of 0.001 and the Adam optimizer. The choice of hyperparameters for this architecture (i.e., the number of layers, channels, latent-space dimension, learning rate and batch-size) were manually tuned to obtain the current performance.  Also, each convolutional layer in the autoencoder utilized a ReLU activation function, with the exception of the output layer and the final layer of the encoder. No regularization was used in the process of training this model and approximately 10\% of the total (non-test) data was set aside for the purpose of validation (i.e., for preventing overfitting through an early-stopping criterion).

\subsubsection{LSTM}

In this section, we introduce architectural details of the LSTM used to advance latent space representations obtained by the CAE for the Burgers problem. We shall be outlining results from two different LSTM architectures: one that is valid for only one choice of $\nu$ and one that is valid for parametric interpolation. We observe that, in general, the latter requires more complex models.

Our basic LSTM architecture for this test case is two cells that are stacked on top of a windowed input of latent space representations. This leads to a windowed-input advancement of dynamics with the output being the prediction of the latent space representation at the next time step. This prediction is then fed back into the framework in an autoregressive manner. Our learning rate for the LSTM is the default 0.001, and we use the Adam optimizer for training. As in the case of the CAE, our cost function is the mean-squared error between predictions and targets. The LSTM hidden cells contain 20 neurons, and the batch size is set to 64 samples. Like in the case of the CAE, we do not employ any regularization and 10\% of the snapshot data is set aside for the purpose of validation. The trained LSTM is deployed for emulating the evolution of the same data in a recursive fashion (i.e., outputs from the LSTM are used as inputs at the next time step).

\subsubsection{CAE-LSTM modeling}

We assess the proposed framework on multiple datasets, each with a single value of $\nu$. Solution fields that vary in time are generated using the analytical solution described in Equation \ref{Burgers_Sol}. In this set of tests, we check the accuracy for different physics ranging from more dissipative at high values of viscosity to more advective at lower values. Error metrics and latent space visualization are provided to evaluate if any trends emerge that generalize in physics. We select four values of $Re=1000,2000,3000,4000$, each with 400 snapshots of the solution field uniformly distributed in time. For the purpose of comparison we also provide results from the POD-Galerkin projection methodology. 
   
\begin{figure}
    \centering
    \includegraphics[width=0.32\textwidth]{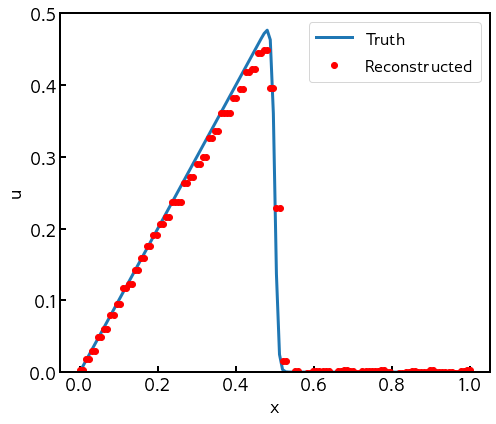}
    \includegraphics[width=0.32\textwidth]{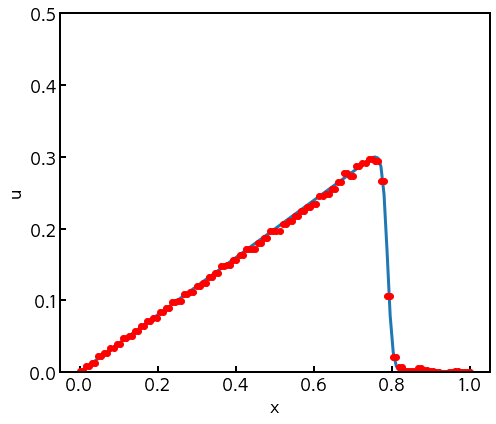}
    \includegraphics[width=0.32\textwidth]{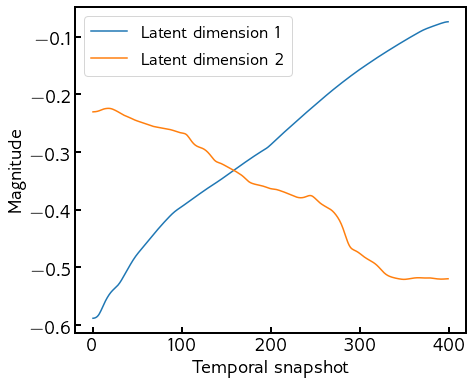}
    \caption{Reconstruction ability of the CAE for initial condition (left) and the final field (middle). Evolution of the latent space (right) for $Re=1000$.}
    \label{Burgers_1000_Rec}
\end{figure}

\begin{figure}
    \centering
    \includegraphics[width=0.32\textwidth]{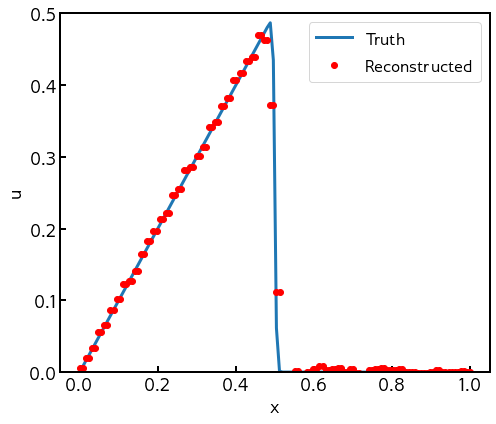}
    \includegraphics[width=0.32\textwidth]{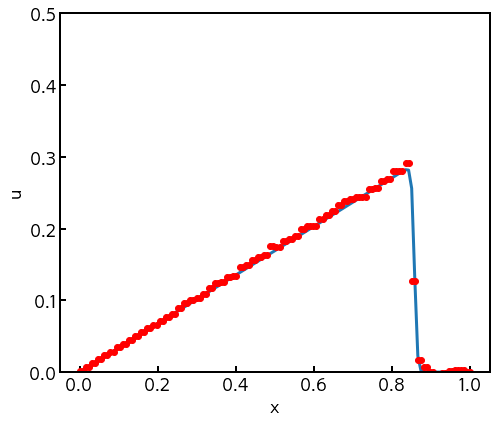}
    \includegraphics[width=0.32\textwidth]{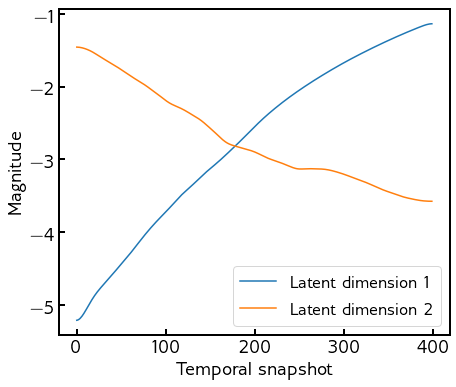}
    \caption{Reconstruction ability of the CAE for initial condition (left) and the final field (middle). Evolution of the latent space (right) for $Re=2000$.}
    \label{Burgers_2000_Rec}
\end{figure}

\begin{figure}
    \centering
    \includegraphics[width=0.32\textwidth]{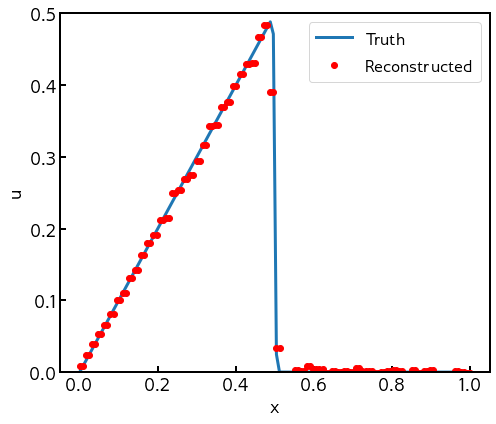}
    \includegraphics[width=0.32\textwidth]{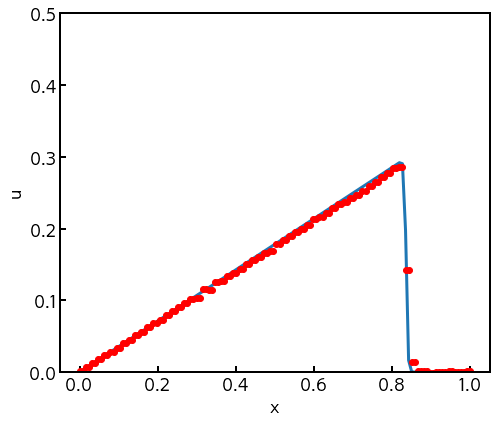}
    \includegraphics[width=0.32\textwidth]{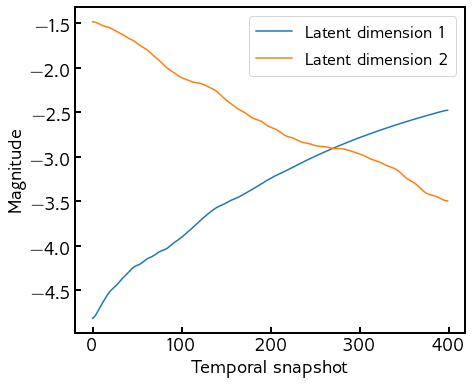}
    \caption{Reconstruction ability of the CAE for initial condition (left) and the final field (middle). Evolution of the latent space (right) for $Re=3000$.}
    \label{Burgers_3000_Rec}
\end{figure}

\begin{figure}
    \centering
    \includegraphics[width=0.32\textwidth]{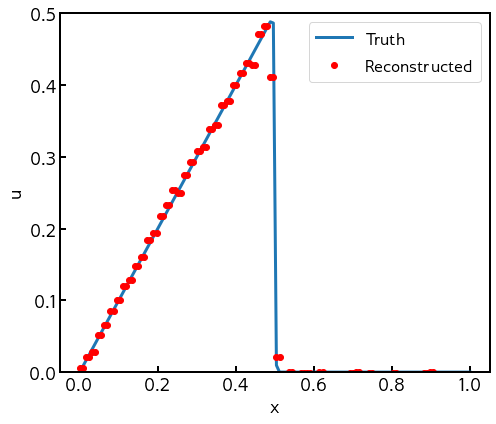}
    \includegraphics[width=0.32\textwidth]{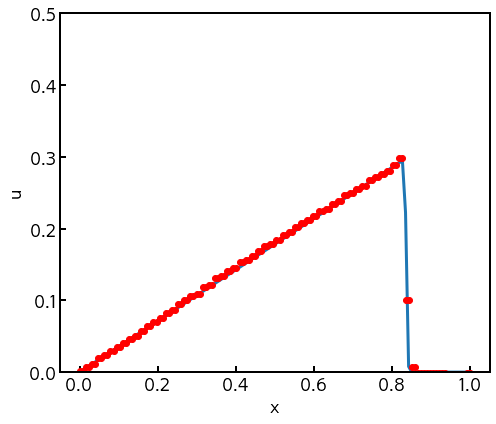}
    \includegraphics[width=0.32\textwidth]{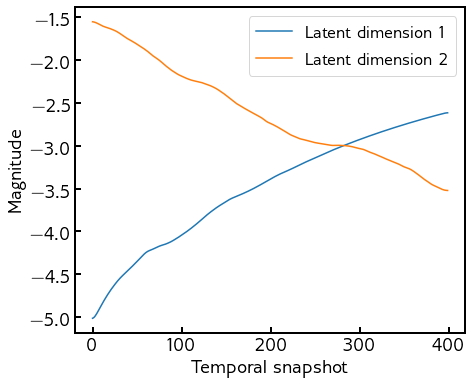}
    \caption{Reconstruction ability of the CAE for initial condition (left) and the final field (middle). Evolution of the latent space (right) for $Re=4000$.}
    \label{Burgers_4000_Rec}
\end{figure}

Figure \ref{Burgers_1000_Rec} shows the performance of CAE deployment for $Re=1000$. This parameter choice leads to viscous effects damping the shock profile as it is advected in the positive $x$ direction. The latent space consisting of two variables has a consistent trend in time which is repeated for other parameters. We draw attention to the difference in magnitude of the latent space variables at the final snapshot. Empirically, this difference is correlated with the dominance of advection to dissipation in the physics of the Burgers equation. Figure \ref{Burgers_2000_Rec} shows similar results for a higher value of $Re=2000$. We remark here that the training for this particular case was completely independent of the other values of $Re$. A good performance in capturing the (now sharper) shock profile is observed. The profile of the latent space evolution is very similar to the previous test case, although the magnitudes of the representation seem to be different. This could possibly be due to scaling through the bias terms of the CAE. Results in Figure \ref{Burgers_3000_Rec} for $Re=3000$ and Figure \ref{Burgers_4000_Rec} for $Re=4000$ show similar behavior in latent space trends as well, indicating that there may be a universality in the compressed representation of this particular problem. This also has implications for the \emph{generation} of new advection-dissipation profiles. We also observe that final time magnitudes of each dimension of the two-dimensional compressed representations appear to be closer to each other with increasing $Re$ perhaps allowing for some interpretability of the latent space. Recall that the different values of $Re$ selected for assessment essentially control the \emph{sharpness} of the shock and have limited effect on the location of the shock. A thorough investigation of interpretability, however, is beyond the scope of this article.  At this point, we have not deployed any latent space model and these assessments are purely related to the  CAE. In the following, we incorporate a latent space time-series model to obtain a 2 degree-of-freedom dynamical model of the advecting shock profile. 

\begin{figure}
    \centering
    \mbox{
    \subfigure[$Re=1000$]{\includegraphics[width=0.48\textwidth]{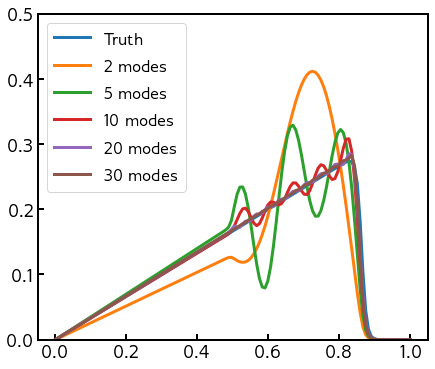}}
    \subfigure[$Re=2000$]{\includegraphics[width=0.48\textwidth]{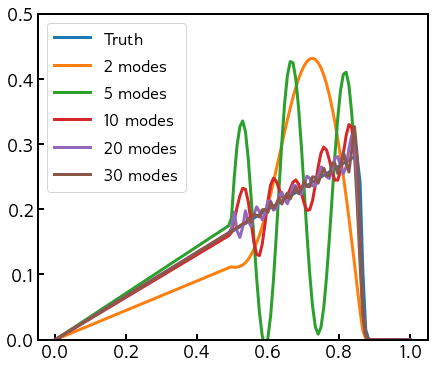}}
    }
    \\
    \mbox{
    \subfigure[$Re=3000$]{\includegraphics[width=0.48\textwidth]{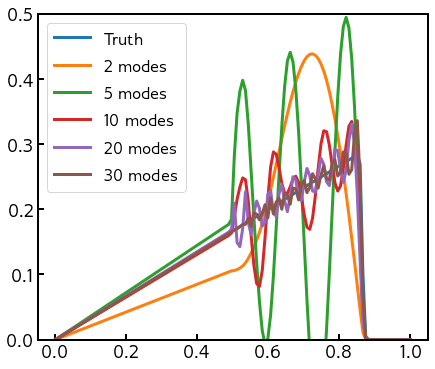}}
    \subfigure[$Re=4000$]{\includegraphics[width=0.48\textwidth]{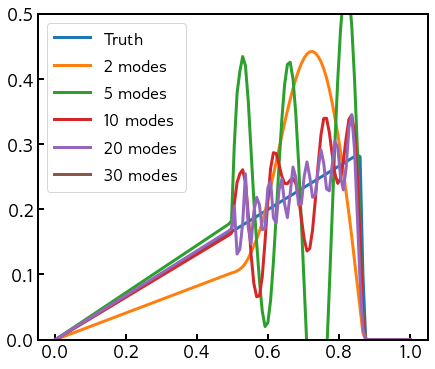}}
    }
    \caption{A demonstration of the limitations of the POD-Galerkin methods for building surrogates of advection dominated partial differential equations. Convergence to the true solution is slow and often limited by numerical instability.}
    \label{POD_GP_Limitation}
\end{figure}

\begin{figure}
    \centering
    \mbox{
    \subfigure[$t=0.5$]{\includegraphics[width=0.48\textwidth]{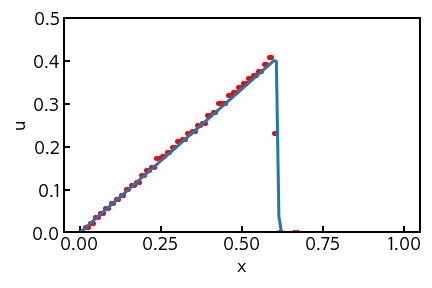}}
    \subfigure[$t=1.0$]{\includegraphics[width=0.48\textwidth]{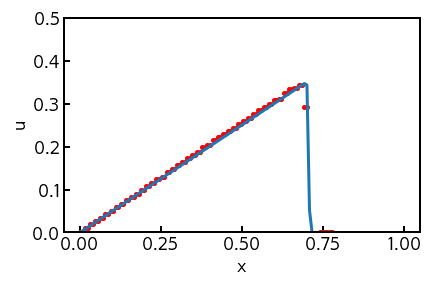}}
    }
    \\
    \mbox{
    \subfigure[$t=1.5$]{\includegraphics[width=0.48\textwidth]{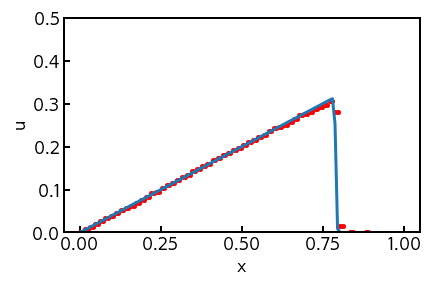}}
    \subfigure[$t=2.0$]{\includegraphics[width=0.48\textwidth]{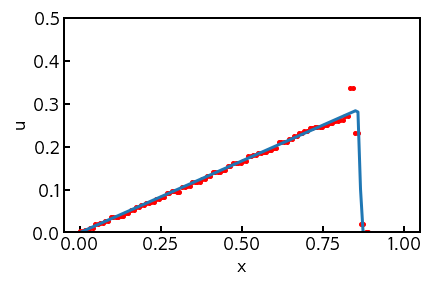}}
    }
    \caption{Reduced-order modeling capability of the CAE for $Re=4000$ showing evolution in physical space. We remind the reader that the system evolution has been performed using an LSTM in latent space, and these images are reconstructed from two degrees of freedom representations.}
    \label{Burgers_4000_ROM}
\end{figure}

\begin{figure}
    \centering
    \includegraphics[width=\textwidth]{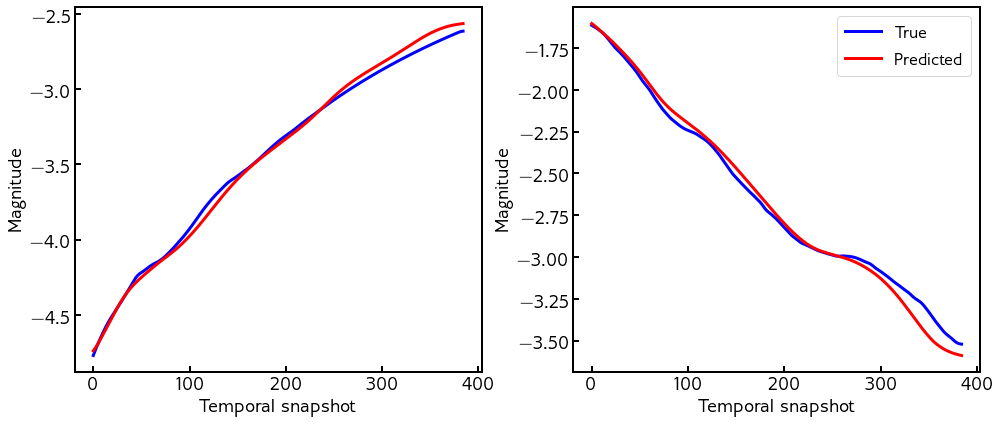}
    \caption{Learning dynamics in latent space obtained using CAE for $Re=4000$. The $y$-axes indicate the magnitudes of the first (left) and second (right) latent space encoding.}
    \label{Burgers_4000_LSTM}
\end{figure}

We now assess the ability of the proposed framework to mimic a standard reduced-order model. We start with an assessment of the POD-GP implementations at different values of $Re$ as shown in Figure \ref{POD_GP_Limitation}. The linear encoding leads to slow convergence of ROM representations to the shock profile. In addition, we also observe high frequency instabilities as the number of retained POD modes is increased for higher values of $Re$. This is due to the use of schemes which are not shock capturing, which causes Gibbs oscillations near the advecting discontinuity. This manifests itself in a solution that fails to converge at $Re=4000$ for 30 retained modes and highlights a critical issue with the reduced-order modeling of advection-dominated problems. Each POD-GP deployment utilized basis vectors from their respective full-order models. In comparison, we show results from the CAE-LSTM implementation in Figure \ref{Burgers_4000_ROM} which shows the ability of the proposed framework to capture the sharp profile advection with only two degrees of freedom. Figure \ref{Burgers_4000_LSTM} shows the prediction of the latent-space model in comparison to the latent space representation by compressing each of the true snapshots. The evolution in the encoded space is recursive, in that the outputs of the LSTM are fed back into the input layer through a windowed input to obtain single time step output. The window is initialized with the true values of the first 10 time steps which implies that, in practice, a short duration of the simulation must be deployed with a full-order model following which the CAE-LSTM can take over non-intrusively. Research is underway to bypass this limitation by appending ghost-points in time to the training data in latent space to mimic a \emph{burn-in} for the windowed input.

\begin{figure}
    \centering
    \includegraphics[width=0.8\textwidth]{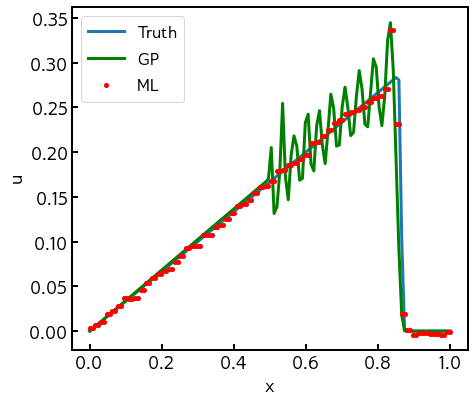}
    \caption{A direct comparison of the POD-GP and CAE-LSTM methods for $Re=4000$ where one can observe noise in the ML predictions even if oscillations are stabilized.}
    \label{GP_CAE_Plot}
\end{figure}

We perform assessments for the CAE-LSTM (as outlined for $Re=4000$ above) for other parameter choices and report error metrics (given by $L_2$-norms at the final time step) in Table \ref{Table1}. These show the accuracy of the framework when compared to the POD-GP method for different POD mode retentions. In general, when dynamics are more advective, the CAE-LSTM has lower errors due to the self-similarity in the advecting shock profile. In comparison, the POD-GP method shows an order of magnitude greater errors at a comparable compression of 2 modes and displays trouble in dealing with strong advective physics for $Re=4000$. Also, the CAE-LSTM, while unable to match POD-GP accuracies at greater mode retentions and lower $Re$, obtains an order of magnitude lower error across different $Re$ for the same latent space dimensions (two degrees of freedom only). This establishes, empirically, that advective physics benefits from nonlinear encoding in space and nonlinear modeling in time for effective surrogates. Table \ref{Table1} is complementary to Figure \ref{POD_GP_Limitation}. While POD-GP shows greater oscillations even at high modal coefficient retention, overall $L_2$-error metrics are comparable (if not better) to the proposed framework. This is elaborated in Figure \ref{GP_CAE_Plot}, which shows that the CAE results in noise in the reconstructed fields even if the oscillations due to the POD-GP implementation are stabilized.

\begin{table}[]
\centering
\begin{tabular}{|c|c|c|c|c|}
\hline
            & Re = 1000 & Re = 2000 & Re = 3000 & Re = 4000 \\ \hline
GP 2 modes  & 4.197e-3  & 5.558e-3  & 6.12e-3   & 6.418e-3  \\ \hline
GP 5 modes  & 1.57e-3   & 7e-3      & 1.244e-2  & 1.65e-2   \\ \hline
GP 10 modes & 1.497e-4  & 5.047e-4  & 1.063e-3  & 1.525e-3  \\ \hline
GP 20 modes & 4.607e-5  & 1.679e-4  & 4.099e-4  & 7.336e-4  \\ \hline
GP 30 modes & 4.938e-5  & 1.102e-4  & 8.333e-5  & NaN       \\ \hline
CAE LSTM    & 4.181e-4  & 3.912e-4  & 1.409e-4  & 1.551e-4 \\
\hline
\end{tabular}
\caption{$L_2-norm$ error metrics for the final time reconstructions of the CAE-LSTM compared against POD-GP. This table outlines results where the CAE-LSTM and POD-GP deployments are trained anew for each $Re$. The CAE-LSTM error is lower for comparable compression (two degrees of freedom).}
\label{Table1}
\end{table}

We now extend the CAE-LSTM for parametric interpolation. By training the framework for full-order datasets generated for different $Re$, our framework can interpolate in a physical regime for quick generation of full-order dynamics at novel parameter choices. We achieve this by appending another scalar component, the viscosity, to the latent space dimension. For training, we obtain snapshots from 19 simulations (i.e., with uniformly varying values of $Re$) and train a common CAE for all of the simulations. This lets us obtain a sequence of latent space representations for each full-order model concatenated with their respective viscosities. We then train an LSTM, also common across all of the simulations, on these sequences in the same manner as the previous experiments. Inferences can then be performed at a novel parameter choice with ease. 

\begin{figure}
    \centering
    \mbox{
    \subfigure[$Re=250$]{\includegraphics[width=0.42\textwidth]{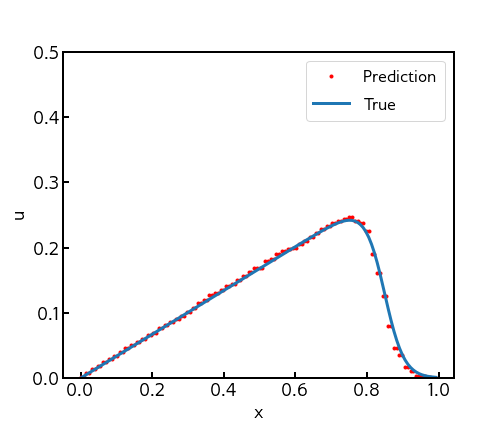}}
    \subfigure[$Re=450$]{\includegraphics[width=0.42\textwidth]{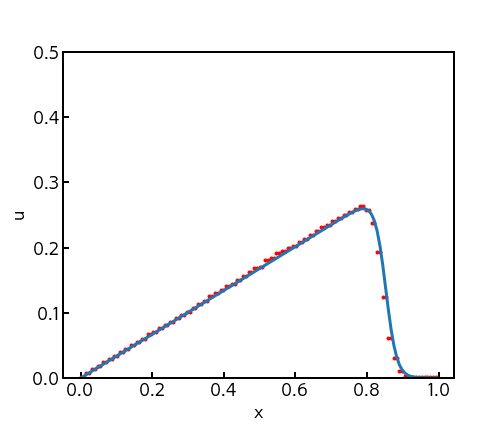}}
    } \\
    \mbox{
    \subfigure[$Re=650$]{\includegraphics[width=0.42\textwidth]{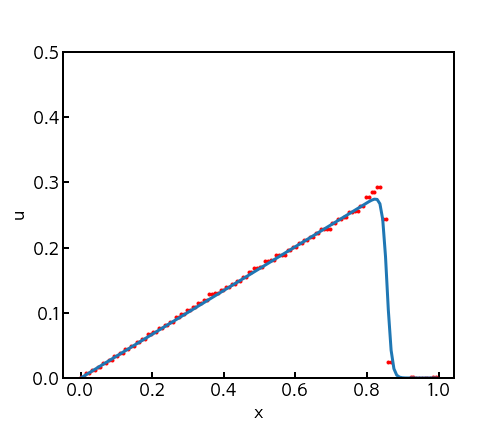}}
    \subfigure[$Re=850$]{\includegraphics[width=0.42\textwidth]{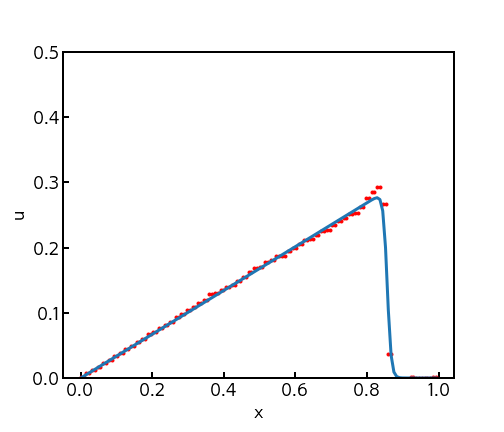}}
    } \\
    \mbox{
    \subfigure[$Re=1050$]{\includegraphics[width=0.42\textwidth]{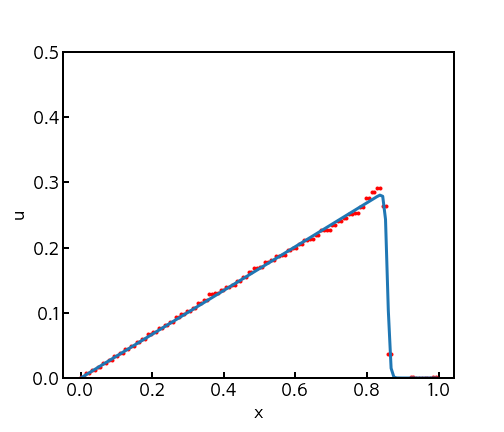}}
    \subfigure[$Re=1250$]{\includegraphics[width=0.42\textwidth]{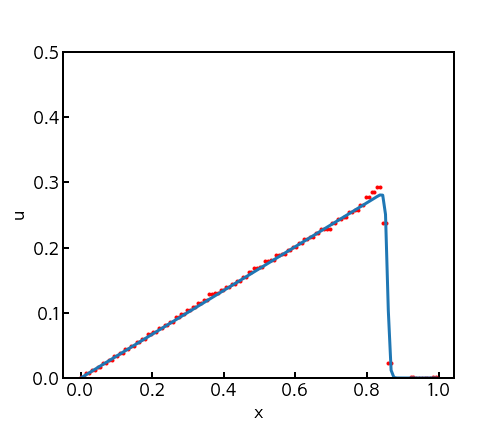}}
    }
    \caption{The ability of a CAE to reconstruct fields sampled from different parameters (Reynolds numbers) showing different sharpness in shock profiles. These snapshots are for parameters that were not included in the training dataset and are obtained by evolving only in the latent space.}
    \label{MP_Reconstruction_Burgers}
\end{figure}

\begin{figure}
    \centering
    \includegraphics[width=0.95\textwidth]{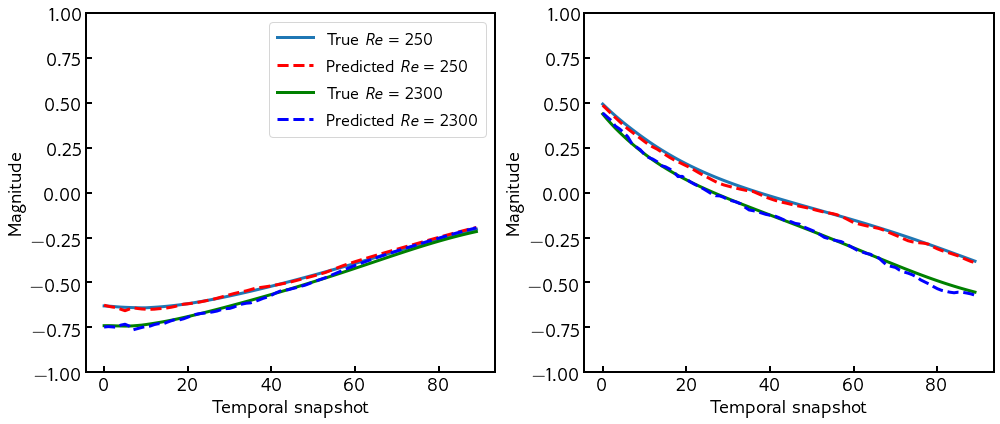}
    \caption{The ability for the parametric LSTM to learn latent space trends for different parameters that are not a part of the training data set. The $y$-axes indicate the magnitudes of the latent space encoding.}
    \label{MP_Burgers_LSTM}
\end{figure}

\begin{figure}
    \centering
    \mbox{
    \subfigure[$t=0.2$]{\includegraphics[width=0.33\textwidth]{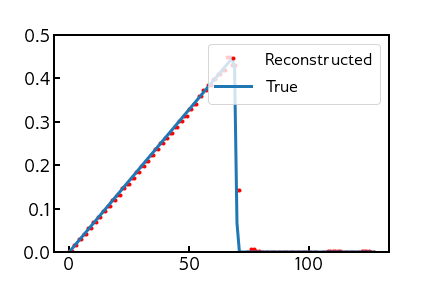}}
    \subfigure[$t=0.4$]{\includegraphics[width=0.33\textwidth]{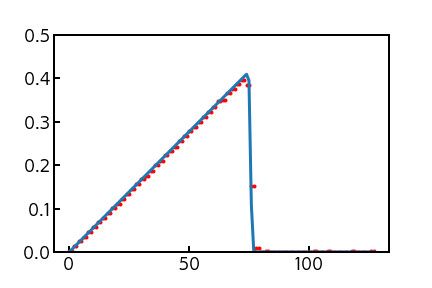}}
    \subfigure[$t=0.6$]{\includegraphics[width=0.33\textwidth]{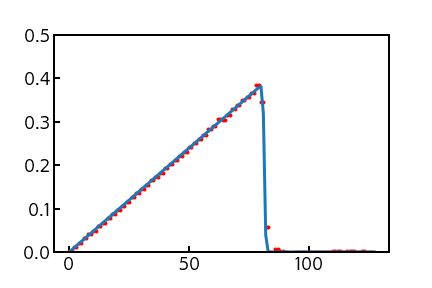}}
    } \\
    \mbox{
    \subfigure[$t=0.8$]{\includegraphics[width=0.33\textwidth]{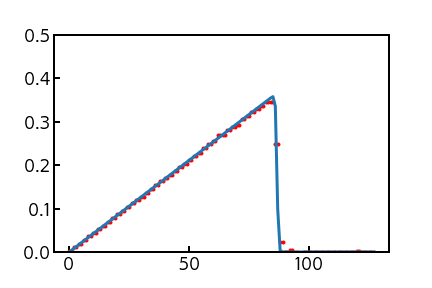}}
    \subfigure[$t=1.0$]{\includegraphics[width=0.33\textwidth]{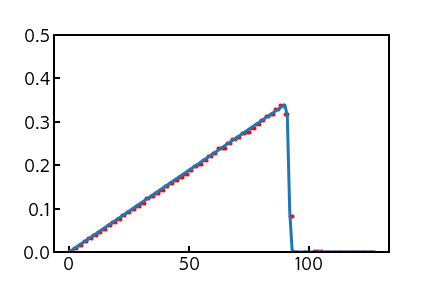}}
    \subfigure[$t=1.2$]{\includegraphics[width=0.33\textwidth]{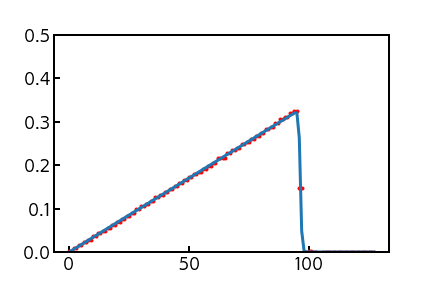}}
    } \\
    \mbox{
    \subfigure[$t=1.4$]{\includegraphics[width=0.33\textwidth]{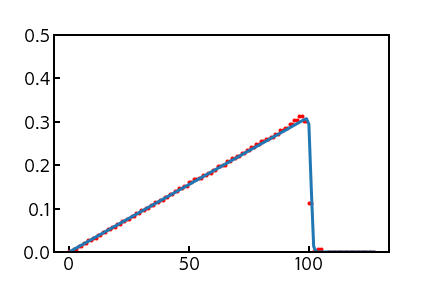}}
    \subfigure[$t=1.6$]{\includegraphics[width=0.33\textwidth]{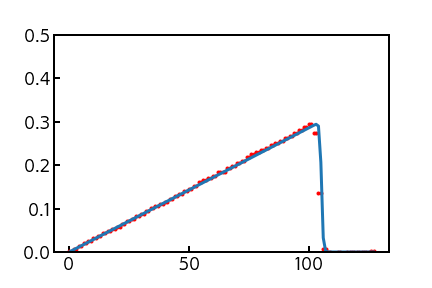}}
    \subfigure[$t=1.8$]{\includegraphics[width=0.33\textwidth]{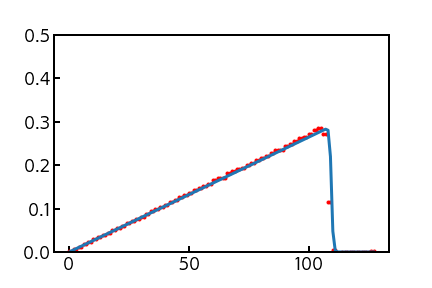}}
    }
    \caption{An example ROM characterized by the CAE-LSTM compared to the full-order solution. This parameter was not a part of the training dataset for either the CAE or the parametric LSTM.}
    \label{Burgers_CAE_LSTM_ROM}
\end{figure}

Our parametric LSTM has a similar architecture to the one we used for single-parameter data in the previous sections. The differences include 40 neurons in the hidden cells and a smaller batch size of 32. We remark that the CAE is identical to the one used previously. The performance of the CAE to reconstruct fields with varying dissipation on the shocked profiles is shown in Figure \ref{MP_Reconstruction_Burgers}. The latent space representation of 2 degrees of freedom is expressive enough to capture the difference in the sharpness of the discontinuity for different viscosities. A parametric LSTM is then trained on these compressed representations with results as shown in Figure \ref{MP_Burgers_LSTM}. We observe that the trends are reproduced appropriately for parameters that were not a part of the training data set. Finally, in Figure \ref{Burgers_CAE_LSTM_ROM}, we demonstrate that reconstructing full-order dynamics for a novel testing parameter accurately adheres to the true solution over time. The final time reconstruction mean-squared errors averaged across different testing viscosities was found to be $1.17e-4$, which was comparable to the cases where training was performed for solely one viscosity.

\subsection{Shallow water equations}
\label{SS:7}

Our two-dimensional assessments utilize the inviscid shallow water equations which are a prototypical system of equations for geophysical flows. The governing equations are hyperbolic in nature and are
\begin{align}
    \begin{gathered}
    \frac{\partial(\rho \eta)}{\partial t}+\frac{\partial(\rho \eta u)}{\partial x}+\frac{\partial(\rho \eta v)}{\partial y} =0, \\
    \frac{\partial(\rho \eta u)}{\partial t}+\frac{\partial}{\partial x}\left(\rho \eta u^{2}+\frac{1}{2} \rho g \eta^{2}\right)+\frac{\partial(\rho \eta u v)}{\partial y} = 0, \\
    \frac{\partial(\rho \eta v)}{\partial t}+\frac{\partial(\rho \eta u v)}{\partial x}+\frac{\partial}{\partial y}\left(\rho \eta v^{2}+\frac{1}{2} \rho g \eta^{2}\right) = 0.
    \end{gathered}
\end{align}
In the above set of equations, $\eta$ corresponds to the total fluid column height, and $(u,v)$ is the fluid's horizontal flow velocity, averaged across the vertical column. Further $g$ is acceleration due to gravity, and $\rho$ is the fluid density, which we fix at 1.0. The first equation captures the law of mass conservation whereas the second two denote the conservation of momentum. Our initial conditions are  
\begin{align}
    \rho \eta (x,y,t=0) &= e^{-\left(\frac{(x-\bar{x})^2}{2(5e+4)^2} + \frac{(y-\bar{y})^2}{2(5e+4)^2}\right)}, \\
    \rho \eta u(x,y,t=0) &= 0, \\
    \rho \eta v(x,y,t=0) &= 0,
\end{align}
while our two-dimensional domain is a square with periodic boundary conditions. We generate data with full-order solves of the above system of equations until $t=0.5$ with a time step of 0.001. Our full-order model uses a 4\textsuperscript{th}-order accurate Runge-Kutta temporal integration scheme and a fifth-order accurate weighted essentially non-oscillatory scheme (WENO) \cite{liu1994weighted} for computing state reconstructions at cell faces. The Rusanov Reimann solver is utilized for flux reconstruction after cell-face quantities are calculated. The reader is directed to \cite{hairer1991solving} for greater discussion of the temporal integration scheme and \cite{maulik2017resolution} for details on WENO and the Riemann solver implementation in two-dimensional problems. For ease of notation we denote $\rho \eta$ as $q_1$, $\rho \eta u$ as $q_2$ and $\rho \eta v$ as $q_3$ in our subsequent discussions. The control parameters in the case of the shallow water equations are $\bar{x}$ and $\bar{y}$, which control the initial location of the Gaussian pulse in the domain. Our goal is to obtain a reduced-basis evolution of a new choice for these control parameters given \emph{a priori} snapshots from full-order forward solves at pre-selected control parameter choices. We use 90 full-order simulations for the training and validation and 10 test simulations for \emph{a posteriori} assessments. One hundred snapshots are utilized for each simulation, i.e., a snapshot is saved every five steps of the time integrator.

\subsubsection{Convolutional autoencoder} \label{Shallow_CAE}

\begin{figure}
    \centering
    \fbox{\includegraphics[trim={2.5cm 8.5cm 2.5cm 2.5cm},clip,width=0.95\textwidth]{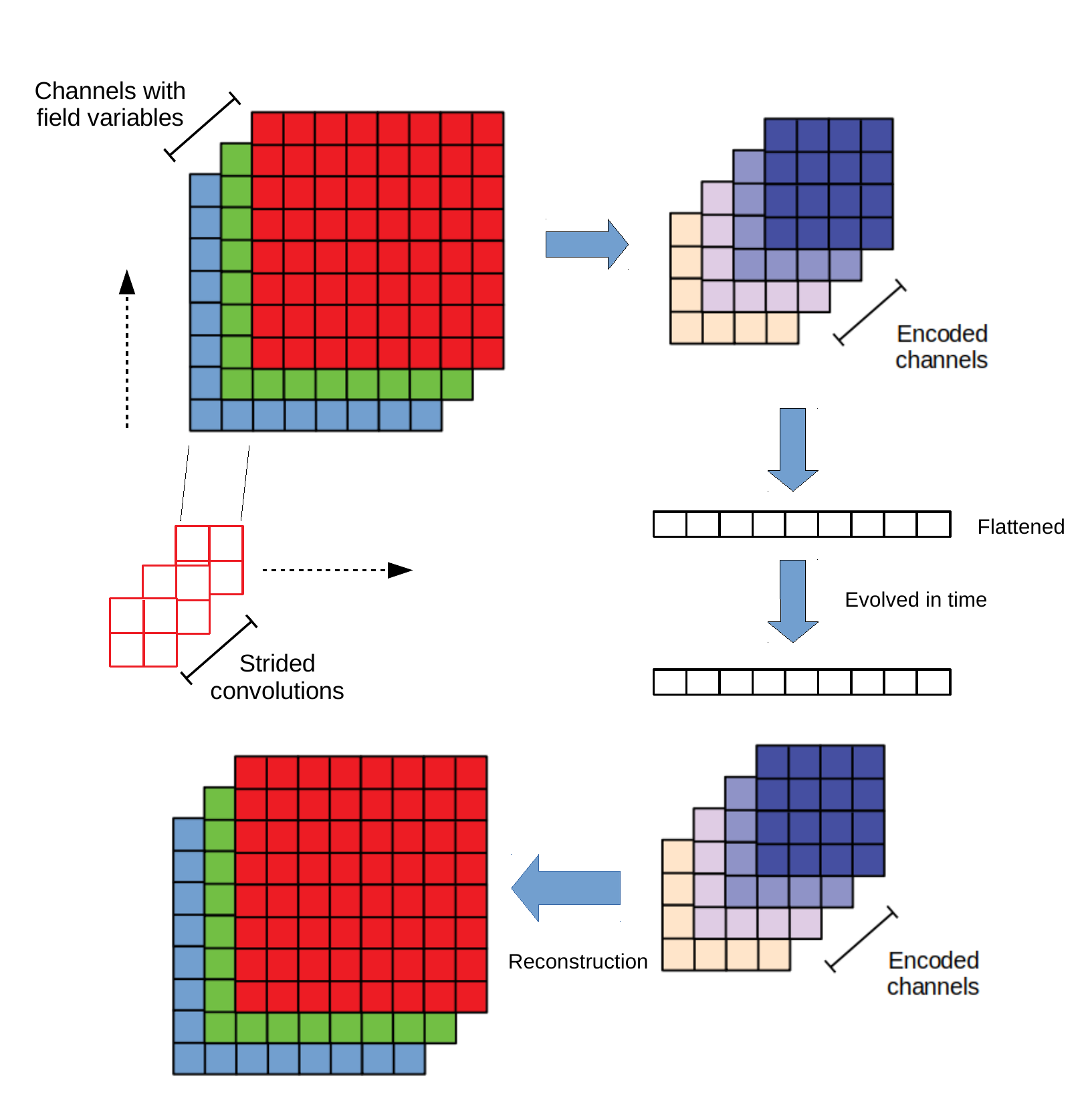}}
    \caption{A schematic of the two-dimensional CAE-LSTM for the shallow water equations. The nonlinear autoencoder embeds the data into latent space, and then the recurrent network can be used for time-series advancement of a flattened representation of the multidimensional system.}
    \label{2D_Schematic}
\end{figure}

\begin{figure}
    \centering
    \includegraphics[trim={0cm 3cm 0 0},clip,width=\textwidth]{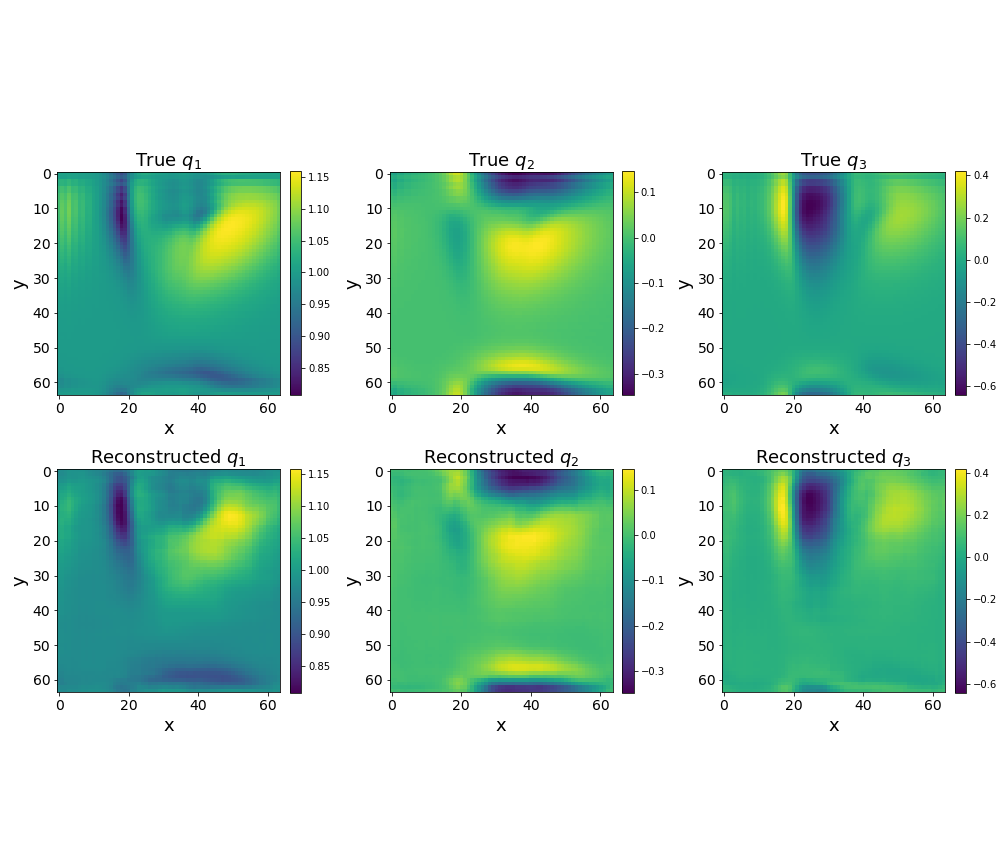}
    \caption{The reconstruction ability of the CAE for the conserved variables in the shallow water equations. This snapshot is from a representative test simulation starting from an unseen initial condition.}
    \label{SWE_CAE_Reconstruction}
\end{figure}

For the nonlinear encoding of the shallow water equations, we use the two-dimensional CAE detailed in the schematic in Figure \ref{2D_Schematic}. Our three conserved variables are encoded using three input and output channels in our autoencoder. We scaled the data to zero mean and unit variance to ensure that losses due to inaccurate reconstruction were weighted fairly across the different variables. We use an architecture that is similar to the Burgers' example in that a bottlenecked framework ensures the compression of the full-order field. A key difference is that the ``bottleneck" layers are supplemented with fully connected layers to allow for an arbitrary latent dimensionality. We choose a latent space of 6 degrees of freedom for this problem which represents an approximate compression ratio of 680. Also, a batch size of 24 with a learning rate of 0.001 was utilized to train the framework. Each two-dimensional convolutional layer and densely connected bottleneck layer utilized the swish activation function, which has been shown to be superior to ReLU for significantly deep networks \cite{ramachandran2017searching}. In contrast, the output layer of the network is a linear layer. 9000 snapshots are randomly partitioned into 8100 for training and 900 for validation, with the latter used for an early stopping criterion. The trained CAE is tested on 1000 snapshots from 10 held-out simulations. Specific details of the architecture such as the number of channels in each layer of the CAE and the size of the pooling may be found in our supporting code. Figure \ref{SWE_CAE_Reconstruction} shows the ability of the decoder to reconstruct from the latent space. 


\subsubsection{LSTM}

We again couple the CAE with an LSTM that is conditioned on the control parameters. In this set of experiments, our control parameter affects the location of a Gaussian pulse applied to $\rho \eta$ at $t=0$. Our goal is to replicate trends of field evolution for a novel initial condition given examples of full-order forward solves to train from. To create training data for the LSTM, we apply the trained CAE to compress the data, then concatenate the parameter information. Our LSTM architecture is 3 cells with 50 neurons in each cell. A batch size of 64 is used with the default learning rate of 0.001 for the Adam optimizer. As outlined previously, 10\% of the total non-test data is set aside for the purpose of validation and early stopping. A time-window of 10 points, utilized for the LSTM forecasts, provided adequately accurate trajectory predictions in the latent space.

\subsubsection{CAE-LSTM modeling}

\begin{figure}
    \centering
    \mbox{
    \subfigure[Latent Dimension 1]{\includegraphics[width=0.42\textwidth]{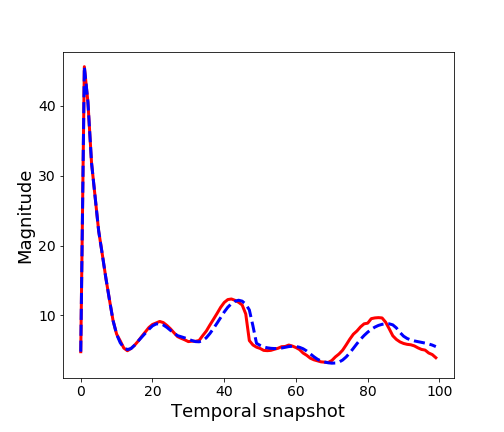}}
    \subfigure[Latent Dimension 2]{\includegraphics[width=0.42\textwidth]{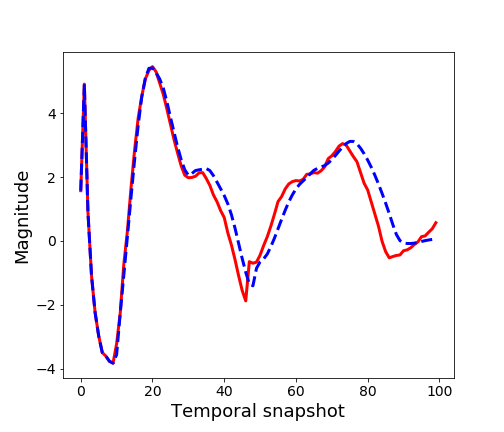}}
    } \\
    \mbox{
    \subfigure[Latent Dimension 3]{\includegraphics[width=0.42\textwidth]{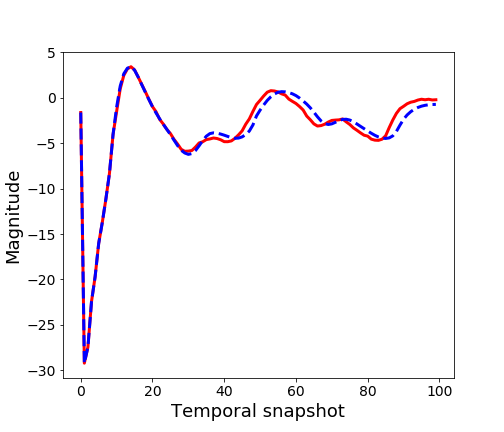}}
    \subfigure[Latent Dimension 4]{\includegraphics[width=0.42\textwidth]{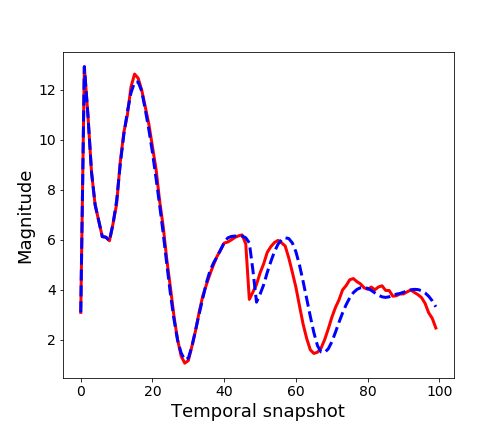}}
    } \\
    \mbox{
    \subfigure[Latent Dimension 5]{\includegraphics[width=0.42\textwidth]{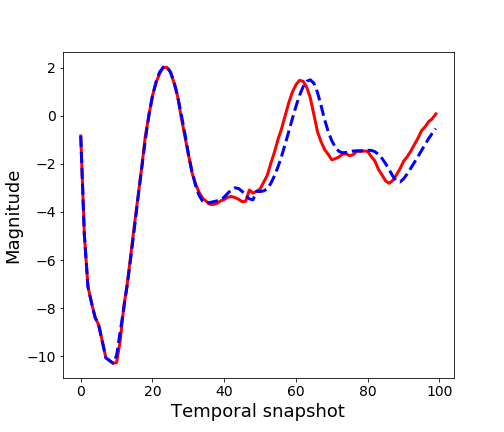}}
    \subfigure[Latent Dimension 6]{\includegraphics[width=0.42\textwidth]{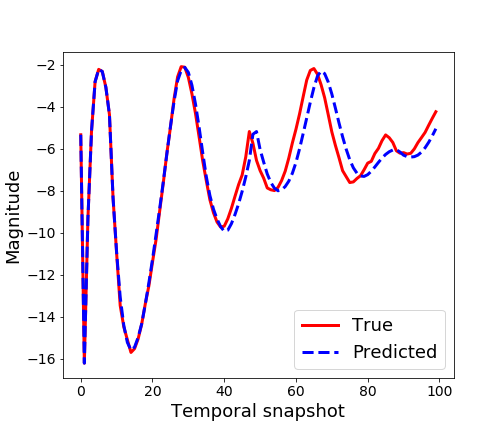}}
    }
    \caption{Hidden space evolution of a testing simulation using a parametric LSTM. The curves, here, indicate the individual degrees of freedom of a 6-dimensional latent space with the $y$-axes indicating their magnitudes.}
    \label{SWE_LSTM_Testing}
\end{figure}

\begin{figure}
    \centering
    \mbox{
    \subfigure[True $q_1$]{\includegraphics[width=0.32\textwidth]{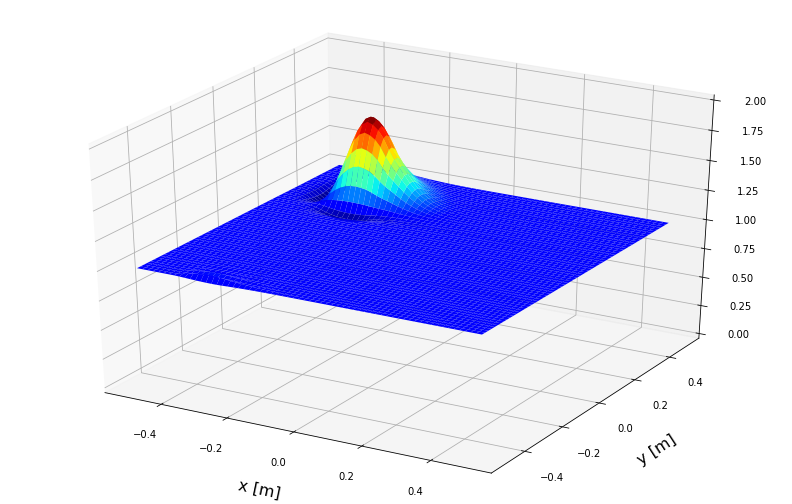}}
    \subfigure[True $q_2$]{\includegraphics[width=0.32\textwidth]{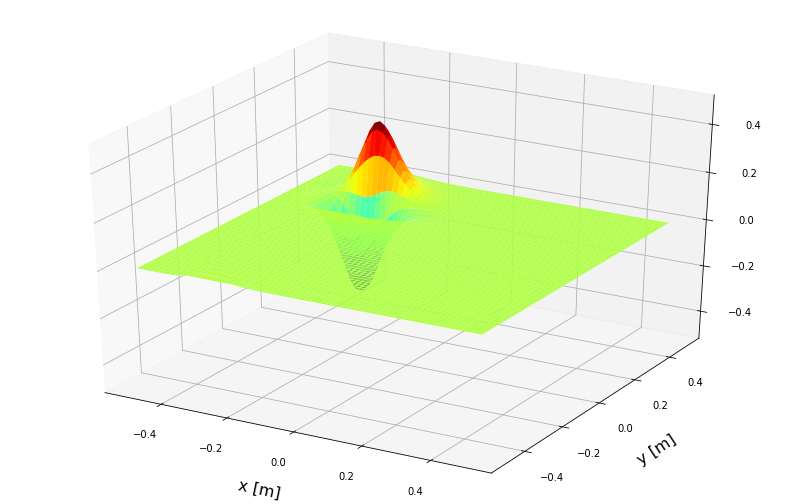}}
    \subfigure[True $q_3$]{\includegraphics[width=0.32\textwidth]{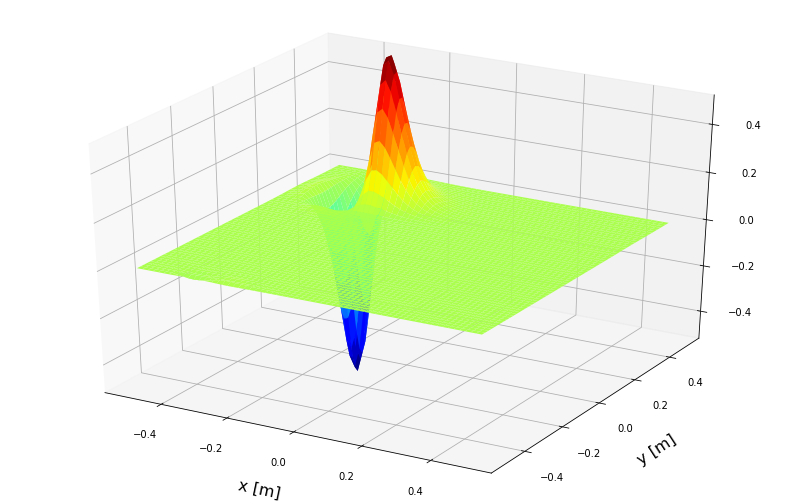}}
    } \\
    \mbox{
    \subfigure[CAE-LSTM]{\includegraphics[width=0.32\textwidth]{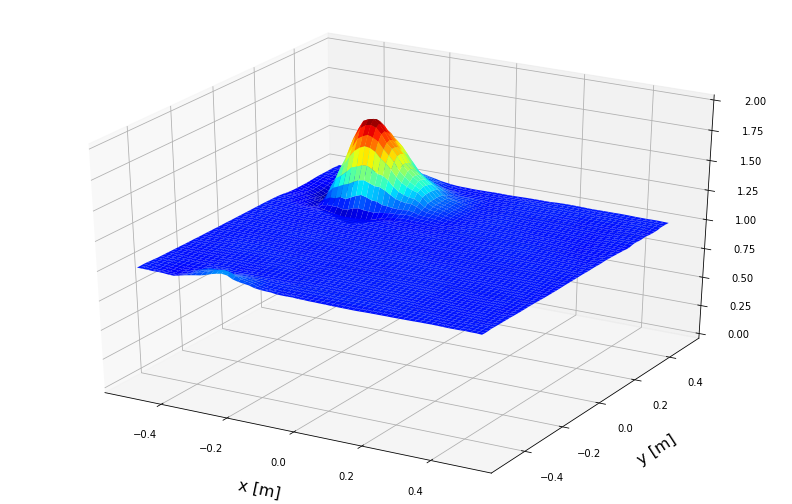}}
    \subfigure[CAE-LSTM]{\includegraphics[width=0.32\textwidth]{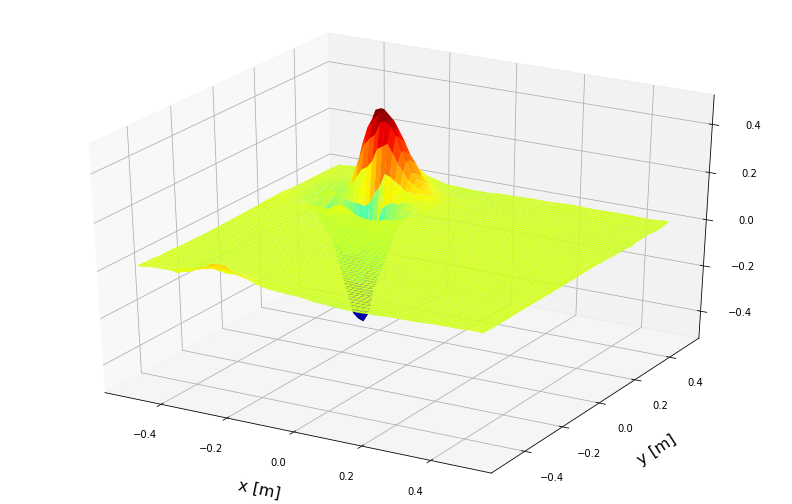}}
    \subfigure[CAE-LSTM]{\includegraphics[width=0.32\textwidth]{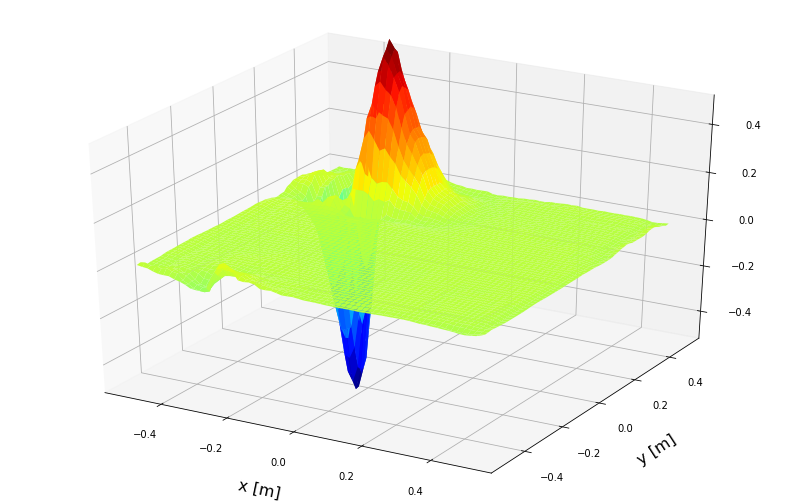}}
    } \\
    \mbox{
    \subfigure[POD-GP (6 modes)]{\includegraphics[width=0.32\textwidth]{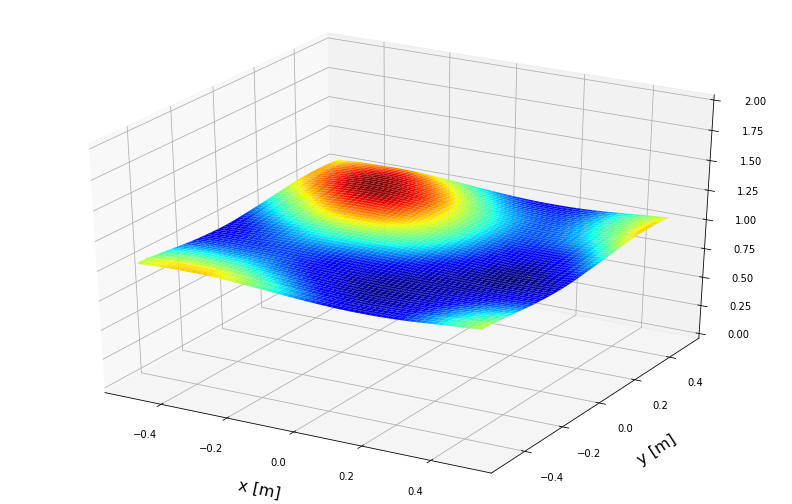}}
    \subfigure[POD-GP (6 modes)]{\includegraphics[width=0.32\textwidth]{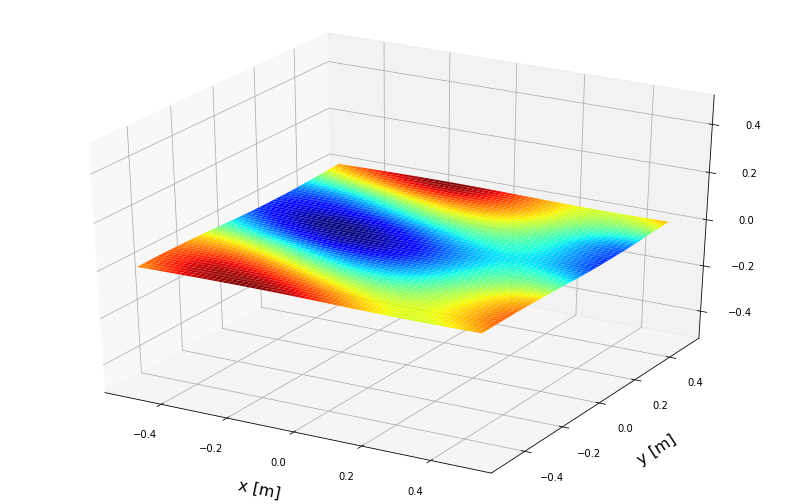}}
    \subfigure[POD-GP (6 modes)]{\includegraphics[width=0.32\textwidth]{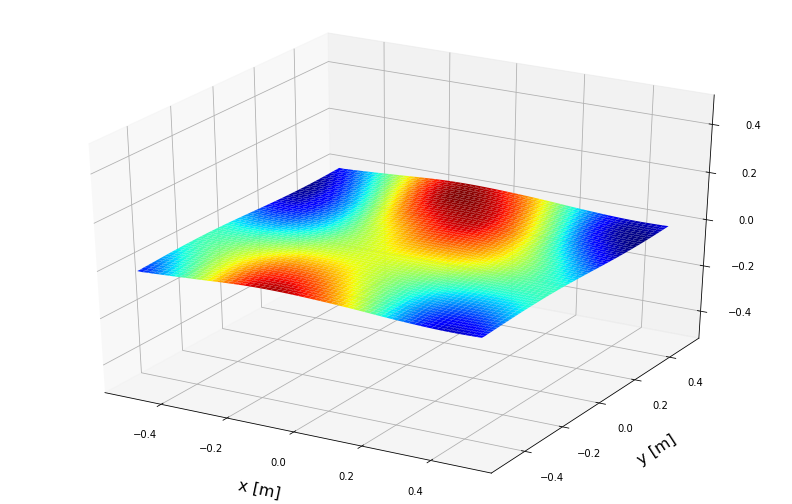}}
    }
    \\
    \mbox{
    \subfigure[POD-GP (40 modes)]{\includegraphics[width=0.32\textwidth]{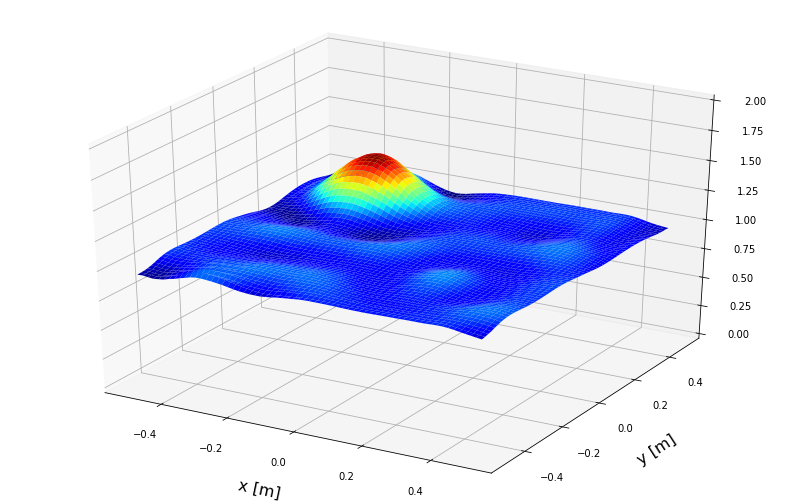}}
    \subfigure[POD-GP (40 modes)]{\includegraphics[width=0.32\textwidth]{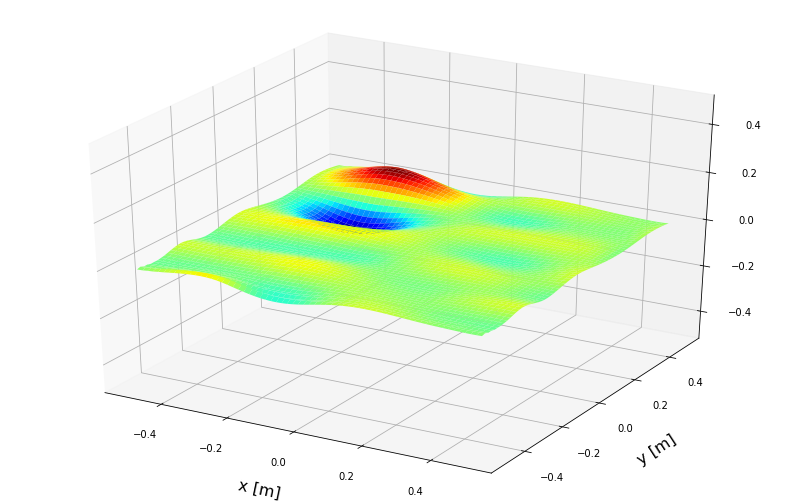}}
    \subfigure[POD-GP (40 modes)]{\includegraphics[width=0.32\textwidth]{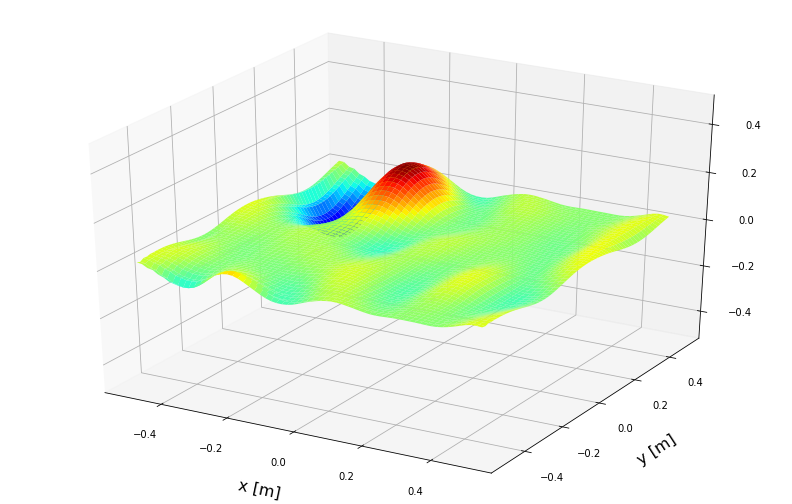}}
    }
    \caption{A qualitative assessment of reconstructed dynamics using the Galerkin projection methodology for a test simulation. The superiority of the CAE reconstruction over POD-GP at the same compression ratio (latent dimension 6) is evident. POD-GP performance improves as we capture more of the variance in the data set by increasing the number of modes.}
    \label{SWE_ROM_1}
\end{figure}

\begin{figure}
    \centering
    \includegraphics[trim={7.5cm 0cm 0 0},clip,width=\textwidth]{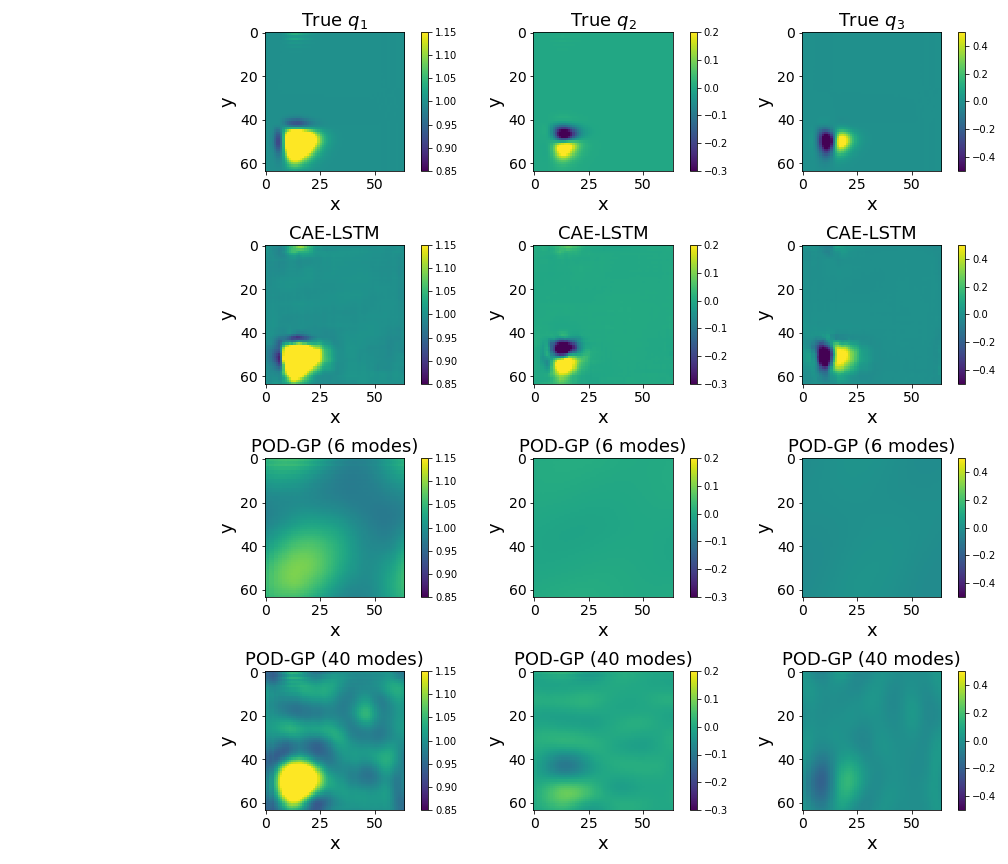}
    \caption{Contour plots showing true, CAE-LSTM and GP obtained results for the three conserved variables at time $t=0.005$. This corresponds to one quarter of the simulation completed. The CAE-LSTM is seen to capture full-order spatial structures accurately in comparison to the POD-GP method (at 6 latent space dimensions).}
    \label{SWE_Contours_1}
\end{figure}

\begin{figure}
    \centering
    \includegraphics[trim={7.5cm 0cm 0 0},clip,width=\textwidth]{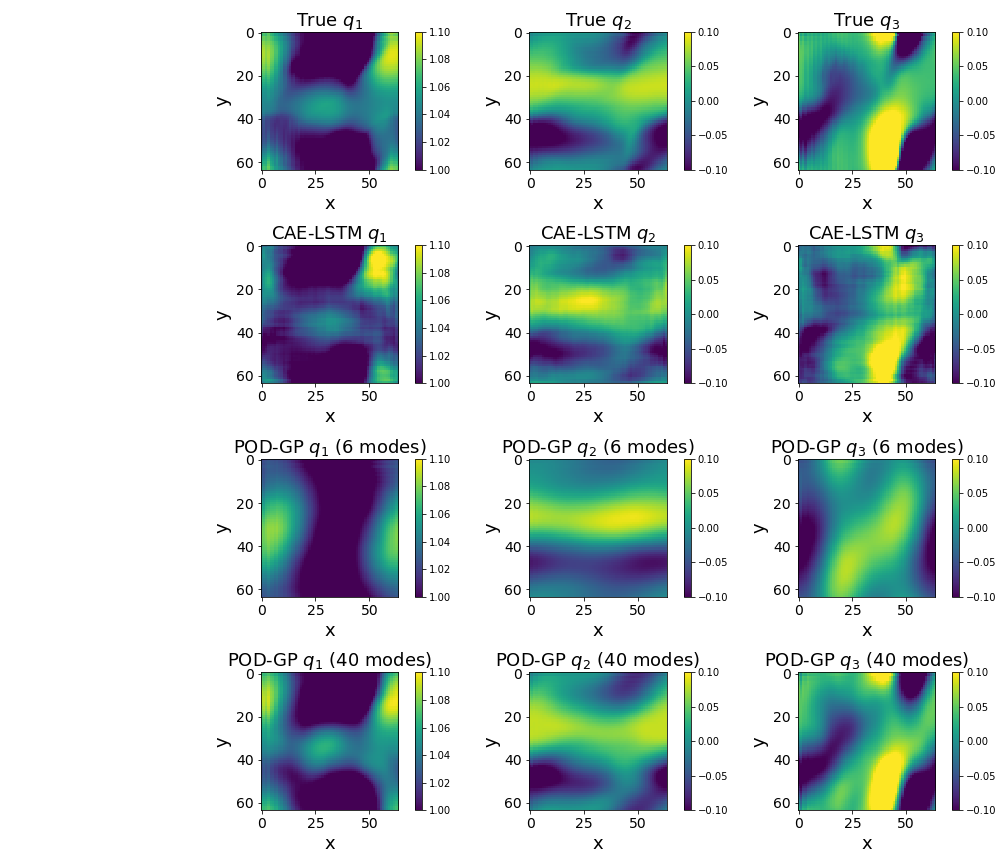}
    \caption{Contour plots showing true, CAE-LSTM and GP obtained results for the three conserved variables at time $t=0.15$. This corresponds to 30\% of the simulation completed. The CAE-LSTM is seen to capture full-order spatial structures accurately in comparison to the POD-GP method (at 6 latent space dimensions).}
    \label{SWE_Contours_2}
\end{figure}

Figure \ref{SWE_LSTM_Testing} shows the ability of the LSTM module to reconstruct dynamical trends in the latent space for a sample test simulation. The reference truth for these curves has been obtained by reconstructing (with use of the CAE) full-order solutions for a test control parameter that was not utilized during training. One can observe that dynamical trends are replicated by the parameterized LSTM. Evolutionary trends towards the end of the dynamics suggest that the dissipation of energy in the system by the numerical method is captured adequately. Figure \ref{SWE_ROM_1} shows the ability of the CAE-LSTM surrogate model to identify coherent spatial features in a sample test simulation. For comparison, we show results from benchmark POD-GP deployments for 6 and 40 retained modes. At an equivalent compression ratio, the CAE-LSTM is able to represent the solution well. At 40 retained modes, the severe truncation of dynamics in POD space still leads to Gibbs' phenomena by POD-GP, which demonstrates the robustness of our proposed method. Contour plots at two representative times are shown in Figure \ref{SWE_Contours_1} and \ref{SWE_Contours_2} where one can clearly observe that the coherent structures in the flow fields are adequately recovered by the CAE-LSTM in comparison to both 6 and 40 mode POD-GP deployments. However, one can also discern that the POD-GP method gradually converges to the true dynamics with increasing modal retention.

In terms of computational costs, the CAE-LSTM was able to provide an LSTM-based latent space forecast at 1.746 seconds per simulation. Reconstruction from latent space for a 100 snapshot simulation required 0.167 seconds. In comparison a POD-GP ROM deployment (using either 6 or 40 retained modes) required an average 24.67 seconds per simulation. The primary cost in POD-GP deployments is the reconstruction of the nonlinear term for the numerical calculation of fluxes which is independent of the number of latent degrees of freedom. The nonlinear term computation for this test case was performed using a fifth-order WENO scheme just like its full-order counterpart and thus is a memory and compute cost that the machine learned model bypasses. In terms of quantitative error metrics, the $q_1$ mean-squared error for all the testing data was 4.8e-4 for CAE-LSTM, 5.6e-4 for POD-GP (6 modes) and 1.7e-4 for POD-GP (40 modes). Similar trends were observed for $q_2$ (4.8e-4, 7.8e-4, 2.6e-4) and $q_3$ (3e-3, 3.3e-3, 1.1e-3). Although the mean-squared error metrics support superiority for POD-GP at 40 retained modes, coherent structure reproduction is more accurate via the CAE-LSTM as demonstrated in the contour plots above. Mean-squared error metrics were affected by the greater amount of fine-scale noise in CAE-LSTM reconstructions. A possible avenue for addressing this limitation is to use intelligent loss functions or embed physics-inspired regularization in the optimization problem.

\section{Discussion and Conclusions}
\label{S:8}

In this study, we propose using a recurrent CAE framework for the reduced-order modeling of systems that are inherently advective and, therefore, high-dimensional. These systems suffer from slow convergence and instability in a linear reduced-basis space given by the POD and a Galerkin projection of the governing equations onto this space. In contrast, we demonstrate that the nonlinear embedding obtained by the CAE and the equation-free dynamics characterization by the LSTM network leads to stable reconstructions of high-dimensional physics in both space and time. We extend our machine learning framework to a parametric formulation where we concatenate the low-dimensional embedding with control parameter information to interpolate between full-order sample points in the data generation phase. Our results indicate that the proposed framework can be used for rapid exploration of a design space conditioned on a set of control parameters. Our framework utilizes a \emph{burn-in} period for the LSTM that necessitates a short compute of less than 10\% of the full-order compute. This is necessary to create a windowed input to the LSTM network. Results on test datasets show a good ability to recover physical trends on unseen control parameter choices. We are currently extending the framework by exploring couplings with active learning wherein we adaptively learn control parameters during training in order to characterize parametric variations optimally. In addition, we are also exploring data-augmentation strategies to preclude the initial compute required for the initial LSTM window in latent space. The former will rely on the generation of so-called \emph{ghost} points to serve as a burn-in to the ROM. Some key challenges also include the ability to incorporate unstructured grid information particularly for problems where there is significant anisotropy in spatial field. There is some promising work in this direction using generalized moving least squares methods \cite{trask2019gmls} and point-cloud networks \cite{kashefi2020point}. The final goal will be to incorporate these surrogate models in design frameworks that may utilize derivative-based or derivative-free optimization.

\section*{Acknowledgments}

The authors acknowledge helpful comments from Dr. Sandeep Madireddy and Dr. Arvind Mohan. This material is based upon work supported by the U.S. Department of Energy (DOE), Office of Science, Office of Advanced Scientific Computing Research, under Contract DE-AC02-06CH11357. This research was funded in part and used resources of the Argonne Leadership Computing Facility, which is a DOE Office of Science User Facility supported under Contract DE-AC02-06CH11357. This paper describes objective technical results and analysis. Any subjective views or opinions that might be expressed in the paper do not necessarily represent the views of the U.S. DOE or the United States Government.

\bibliographystyle{elsarticle-num-names}
\bibliography{references.bib}







\end{document}